\documentclass{article} 
\usepackage{iclr2027_conference,times}

\usepackage{amsmath,amsfonts,bm}

\def\eqref#1{equation~\ref{#1}}

\def\1{\bm{1}}

\DeclareMathAlphabet{\mathsfit}{\encodingdefault}{\sfdefault}{m}{sl}
\SetMathAlphabet{\mathsfit}{bold}{\encodingdefault}{\sfdefault}{bx}{n}

\usepackage{graphicx}
\usepackage{amsmath}
\usepackage{booktabs}
\usepackage{pifont}   
\usepackage{amssymb}  
\usepackage{multirow} 
\usepackage{soul}
\usepackage{ifthen}
\usepackage{etoolbox}
\usepackage{subcaption}
\usepackage{tcolorbox}
\newcommand{\gain}[1]{\colorbox{green!20}{#1}}
\newcommand{\loss}[1]{\colorbox{red!20}{#1}}
\usepackage{epigraph}
\usepackage{tcolorbox}
\usepackage{listings}
\usepackage{booktabs}
\usepackage{colortbl}
\usepackage{xcolor}
\definecolor{lightgreen}{RGB}{220,255,220}
\tcbuselibrary{skins}

\newcommand{\cmark}{\textcolor{green!60!black}{\ding{51}}}
\newcommand{\xmark}{\textcolor{red}{\ding{55}}}

\tcbset{
    promptbox/.style={
        colback=red!8!white,
        colframe=red!40!white,
        arc=6pt,
        boxrule=0.8pt,
        left=8pt, right=8pt, top=6pt, bottom=6pt,
        fonttitle=\small\bfseries\color{red!60!black},
        coltitle=red!60!black,
        attach boxed title to top left={yshift=-2mm, xshift=6mm},
        boxed title style={colback=red!8!white, colframe=red!40!white, arc=4pt, boxrule=0.8pt},
        fontupper=\small\ttfamily,
    }
}

\usepackage[utf8]{inputenc} 
\usepackage[T1]{fontenc}    
\usepackage{hyperref}       
\usepackage{url}            
\usepackage{booktabs}       
\usepackage{amsfonts}       
\usepackage{nicefrac}       
\usepackage{microtype}      
\usepackage{xcolor}         
\usepackage{wrapfig}

\definecolor{promptbg}{HTML}{F8F9FA} 
\definecolor{promptframe}{HTML}{DEE2E6} 
\definecolor{prompttext}{HTML}{212529} 

\lstdefinestyle{llmprompt}{
    backgroundcolor=\color{promptbg},
    rulecolor=\color{promptframe},
    basicstyle=\ttfamily\footnotesize\color{prompttext},
    frame=single,                   
    frameround=tttt,                
    breaklines=true,                
    breakatwhitespace=true,         
    columns=fullflexible,           
    keepspaces=true,                
    captionpos=b,                   
    xleftmargin=1em,                
    xrightmargin=1em,
    showstringspaces=false          
}

\title{Trident: How to Break Deep Reinforcement Learning Cyber Defenses (Agentic)}

\iclrfinalcopy
\author{%
Ryozo Masukawa$^{1}$,\\
Ian Bryant$^{1}$,
Armita Kazeminajafabadi$^{2}$,\\
Sanggeon Yun$^{1}$,
Hyunwoo Oh$^{1}$,
SungHeon Jeong$^{1}$,\\
Nathaniel D. Bastian$^{3}$,
Mahdi Imani$^{2}$,\\
Mohsen Imani$^{1}$\\[1ex]
$^{1}$University of California, Irvine\\
$^{2}$Northeastern University\\
$^{3}$Johns Hopkins University\\
\texttt{rmasukaw@uci.edu}
}

\begin{document}

\maketitle
\begin{abstract}
Autonomous cyber defense systems based on Deep Reinforcement Learning (DRL) have attracted significant research attention, yet remain evaluated almost exclusively against static, heuristic red agents, leaving their robustness against adaptive threats critically understudied. Meanwhile, recent advances in Reinforcement Learning with Verifiable Rewards (RLVR) have improved LLM reasoning, but their integration into cybersecurity remains elusive due to the absence of suitable benchmark environments and interaction datasets. To bridge this gap, we introduce \textbf{Trident}, an agentic LLM red teaming framework comprising three components: a dynamic benchmark with isolated sandbox servers spanning CybORG CAGE 4 and CyberWheel, a dataset comprising over 13,000 high-fidelity red-blue interaction trajectories for RLVR, and a ``Code-as-Policy'' RLVR agentic architecture (\textbf{Trident Agentic}). The latter reformulates red agent training as a contextual bandit via a tripartite Log Summarizer--Planner--Coder design, where a trainable Planner generates complete attack strategies from compressed execution logs, which a frozen Coder translates into executable Python policies deployed against live DRL defenders. Empirical evaluations reveal a consistent brittleness in existing defenses: with a single trainable 7B planner degrades blue-agent performance by an average of \textbf{628\%} across four state-of-the-art CAGE 4 defenders and eliminates over \textbf{99\%} of the defender's episodic reward in CyberWheel, while autonomously discovering novel behaviors such as decoy avoidance and adaptive state prioritization that static heuristics entirely fail to uncover. Code is available \href{https://github.com/BiasLabProjects/Trident}{here}\footnote{\href{https://github.com/BiasLabProjects/Trident}{https://github.com/BiasLabProjects/Trident}}.
\end{abstract}


\vspace{-2mm}
\section{Introduction}
\vspace{-2mm}
\label{sec:intro}
Modern computer networks face escalating cyber threats that increasingly outpace manual defense capabilities~\citep{aicyberchallengeOverviewx2013}. This has driven a long line of research in autonomous cyber defense (ACD): from early rule-based expert systems~\citep{neumann1999experience, taxonomy_acd} and anomaly detection~\citep{lin2022bert, zhao2023yet}, to deep reinforcement learning (DRL) agents~\citep{hmarl, cybermonic, victim_marl, castro2025large} that actively monitor network states and recover compromised hosts in real-time. DRL-based defenders, in particular, remain critically understudied against adaptive threats~\citep{kiely2025exploring, cyberwheel}: they are evaluated almost exclusively against static, heuristic red agents with fixed behavioral signatures, leaving their robustness against adaptive attackers unknown. A recent systematization of DRL for cybersecurity raises the same concern, identifying the fixed, non-adaptive attacker as one of the field's most prevalent evaluation pitfalls~\citep{sok_usenix}.
\begin{figure}[t]
    \centering
    \includegraphics[width=\linewidth]{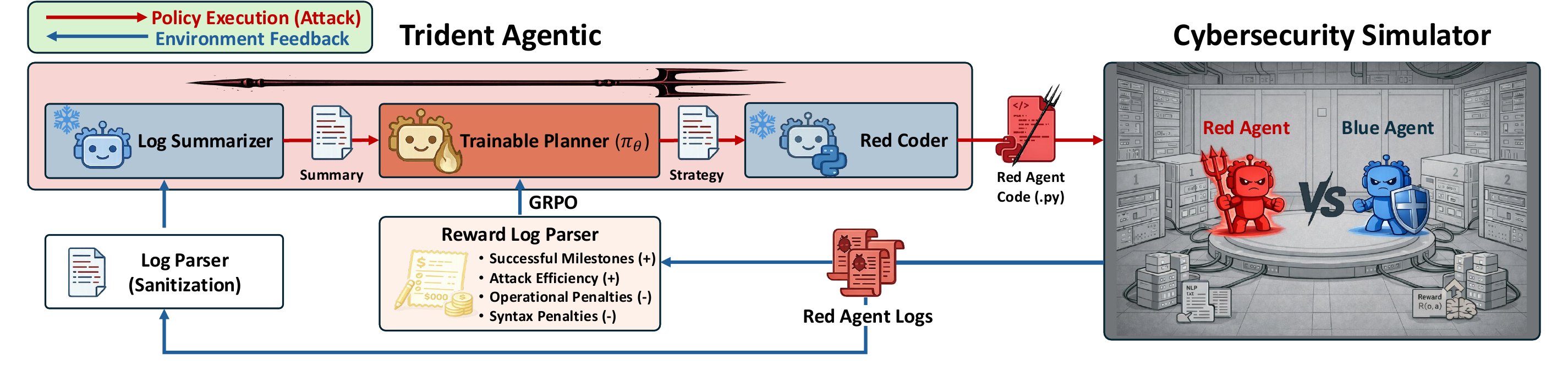}
    \vspace{-2mm}
    \caption{\textbf{Overview of the Trident Agentic framework.} A Log Summarizer observes and summarizes sanitized logs, enabling a trainable Planner ($\pi_\theta$) to formulate abstract tactics. A frozen Coder translates the tactic into an executable Python policy deployed against a live DRL defender. Execution logs provide a verifiable reward signal to update $\pi_\theta$ via GRPO.}
    \label{fig:overview}
    \vspace{-8mm}
\end{figure}

The ideal red agent for rigorous evaluation should be adaptive: capable of observing defensive behavior, reasoning about weaknesses, and synthesizing novel attack strategies. Recent progress in agentic Large Language Models (LLMs) makes this vision increasingly plausible and urgent: Claude Mythos~\citep{anthropicClaudeMythos} has demonstrated that large-scale frontier agentic LLMs can autonomously identify and exploit zero-day vulnerabilities across major operating systems and web browsers. Research on agentic cybersecurity frameworks has made significant advancements across Capture The Flag (CTF) challenges~\citep{abramovich2025enigma, shao2024nyuctfbench, zhang2025cybench}, penetration testing~\citep{deng2023pentestgpt}, and vulnerability reproduction~\citep{zhu2025cvebench, iclr26_cybergym, zhuo2026cyberzero, suryanto2026redsage}. However, all of these works deploy models via prompting or supervised fine-tuning on static datasets, and none train an agent through direct interaction with an active defender using Reinforcement Learning (RL), nor do they provide the infrastructure to do so. 

DeepSeek-R1~\citep{guo2025deepseek} has popularized Reinforcement Learning with Verifiable Rewards (RLVR), where an LLM policy learns to maximize rewards derived from objective, environment-provided rules rather than relying solely on human-labeled examples. RLVR has since shown significant advances across mathematical reasoning~\citep{cobbe2021gsm8k, datagen_evidence3, preliminary1}, software engineering~\citep{wei2025swerl, zeng-etal-2025-acecoder_agent_example2}, information retrieval~\citep{datagen_evidence2}, and computer vision~\citep{visual_rft_agent_example1}. Yet its integration into cybersecurity has remained elusive because execution environments are often ephemeral or restricted~\citep{zhuo2026cyberzero}, even though the latest large-scale studies identify RLVR as a promising direction yet leave its realization entirely unexplored~\citep{yu-etal-2025-primus}. A key reason is the absence of a suitable benchmark: one that provides stable, repeatable adversarial environments, verifiable reward signals, and active defenders against which red agents can be trained and evaluated.

High-fidelity simulators, such as CybORG CAGE 4~\citep{kiely2025exploring} and CyberWheel~\citep{cyberwheel}, offer a concrete path forward. They provide stable, repeatable adversarial environments designed to closely replicate real-world network operations. CAGE 4 models a realistic segmented enterprise network with cooperative multi-agent defenders operating under partial observability, while CyberWheel grounds its attack actions in real-world MITRE ATT\&CK~\citep{mitreMITREATTampCKxAE} techniques via Atomic Red Team~\citep{atomicredteamAtomicTeam}, supporting higher attack realism. Together, they offer realistic network attack and defense dynamics with deterministic, verifiable reward signals from system events, the properties needed to support RLVR training against active defenders. This motivates a critical question: \textit{Can agentic LLMs be trained via RLVR to compromise state-of-the-art DRL defenders?}

To answer this question, we introduce \textbf{Trident} (\autoref{fig:overview}), a dynamic red teaming benchmark spanning CAGE 4 and CyberWheel. Trident provides isolated sandbox servers that execute generated attack policies against live DRL defenders to produce verifiable reward signals, alongside a dataset of over 13,000 high-fidelity red-blue interaction trajectories for RL-based post training, to the best of our knowledge, the first resource to support RLVR training against active DRL defenders.

To demonstrate the utility of this infrastructure, we further introduce \textbf{Trident Agentic}, an agentic LLM framework trained via RLVR under a ``Code-as-Policies''~\citep{code_as_policies, wei2025swerl} paradigm, where an agent's policy is instantiated as executable, LLM-generated code. Rather than selecting actions step-by-step under a Decentralized Partially Observable Markov Decision Process (Dec-POMDP), Trident Agentic reformulates red agent training as a contextual bandit.
A trainable \textit{Planner} receives a compressed summary of past execution logs from a \textit{Log Summarizer} and generates an attack strategy in a single forward pass. A frozen \textit{Coder} then translates this strategy into an executable Python policy.

Using Trident Agentic, we reveal a critical gap in how autonomous cyber defenders are currently assessed. Trident Agentic systematically compromises state-of-the-art DRL blue agents by discovering attack sequences that expose weaknesses in their learned policies, weaknesses that traditional heuristic red agents entirely fail to uncover. On average, it degrades blue-agent returns by \textbf{628\%} across the four CAGE 4 defenders and reduces the CyberWheel defender's reward by over \textbf{99\%} relative to the default heuristic baseline. These results suggest that current autonomous defenders may be more brittle against adaptive attackers than prior evaluations have indicated.
Our primary contributions are as follows:
\begin{itemize}
    \item \textbf{Trident evaluation benchmark and dataset}: We introduce a cross-environment red teaming benchmark spanning CAGE 4 and CyberWheel. This framework utilizes isolated sandbox servers to produce verifiable reward signals against live DRL defenders. We release over 13,000 red-blue interaction trajectories, supporting RLVR training on LLMs.  
    
    \item \textbf{Trident Agentic}: We introduce a tripartite agentic LLM architecture that leverages the ``Code-as-Policies'' paradigm to reformulate red agent training from a Dec-POMDP into a contextual bandit via RLVR, enabling the direct generation of executable attack strategies. 

    \item \textbf{Empirical evidence of a consistent DRL brittleness}: Trident Agentic causes an average \textbf{628\%} performance loss on four state-of-the-art CAGE 4 defenders and a \textbf{99\%} reward collapse in CyberWheel, with a trained 7B SLM surpassing frontier-scale models, revealing that current static-adversary evaluation practices critically underestimate the brittleness of autonomous cyber defenses.
\end{itemize}

Beyond its immediate contributions, Trident's RESTful sandbox architecture, where generated policies are submitted and rewards are returned, enables seamless integration of future, higher-fidelity environments as drop-in servers. This plug-and-play design not only supports the continued development of adaptive red agents, but equally opens the door to training blue agents robust against generative attackers, establishing Trident as an extensible platform for the long-term co-evolution of autonomous cyber offense and defense.
\vspace{-5mm}
\section{Preliminaries}
\vspace{-3mm}

\vspace{-1mm}
\subsection{Adversarial cyber environment as dec-POMDP}
\vspace{-2mm}

Previous studies model the autonomous cyber defense task from the defender's perspective as a Dec-POMDP~\citep{hmarl, cybermonic}, defined by the tuple $\mathcal{M} = (\mathcal{I}, \mathcal{S}, \mathcal{A}, \mathcal{T}, \mathcal{R}, \Omega, \mathcal{O}, \gamma)$. Here, $\mathcal{I}$ represents the team of agents including $M \ge 1$ defending agents. $\mathcal{S}$ is the true network state space, $\mathcal{A}$ is the joint defense action space, and $\mathcal{R}(s, a)$ is the shared reward for the defending team. Each defending agent $i \in \mathcal{I}$ perceives the environment only through its local observation via the observation function $\mathcal{O}(o|s', a)$, reflecting the partial observability inherent to realistic network defense. Crucially, in these formulations, red agents and optionally, benign users are not treated as independent learning agents within $\mathcal{I}$, but rather as fixed logic embedded within the environment. The state transition dynamics are conditioned on the red agent policy code $c \in \mathcal{C}$, written as $\mathcal{T}(s'|s, a, c)$, reflecting that the network state trajectory is jointly determined by the environment dynamics and the static logic embedded in $c$.
In existing benchmarks such as CAGE 4 ($M>1$) and CyberWheel ($M=1$), $c$ is instantiated as a heuristic agent such as a finite state machine, meaning blue agents are evaluated against a non-adaptive adversary\footnote{CyberWheel also supports DRL-based red agents; however, we fix $c$ to a static heuristic in both environments for experimental consistency with CAGE 4.}. This fixity leaves critical vulnerabilities in learned blue policies undetected. 
Since $c$ is executable code that modern LLMs can naturally generate, we propose 
training an LLM meta-policy $\pi_\theta$ to directly synthesize $c$ conditioned 
on execution feedback from prior trajectories.
\vspace{-2mm}
\subsection{Strategic red teaming as a contextual bandit}
\vspace{-2mm}
Directly applying $\pi_\theta$ within the step-by-step POMDP loop is intractable due to prohibitive computational cost~\citep{castro2025large}. To circumvent this, Trident adopts a \textbf{Code-as-Policies}~\citep{code_as_policies} paradigm: rather than sequentially emitting discrete primitive actions, $\pi_\theta$ expresses its entire strategy as executable source code that captures complex control flows and semantic reasoning. Specifically, $\pi_\theta$ receives a context $C$ consisting of environment descriptions and historical feedback, and generates a complete, executable Python policy $c \in \mathcal{C}$ in a single forward pass $c \sim \pi_{\theta}(c \mid C)$.
Since $\pi_\theta$ generates $c$ in a single step and receives a reward only upon full execution of $c$ within $\mathcal{M}$, this naturally reduces to a \textbf{Contextual Bandit}~\citep{bandit1, bandit2, cobbe2021gsm8k, preliminary1} 
problem. 
The training objective is to maximize $J(\theta)$, the expected cumulative return 
when $c$ is executed within $\mathcal{M}$:
\begin{equation}
\label{eq:bandit}
J(\theta) = \mathbb{E}_{c \sim \pi_{\theta}(\cdot|C)} \left[ \mathbb{E}_{\tau \sim \pi_c} \left[ \sum_{t=0}^{T} \mathcal{R}(s_t, a_t) \right] \right]
\end{equation}
The inner expectation $\mathbb{E}_{\tau \sim \pi_c}$ reflects multiple layers of 
stochasticity: even for a deterministic $c$, the resulting trajectory $\tau$ 
remains stochastic due to the non-deterministic transition dynamics $\mathcal{T}$ 
and the probabilistic interventions of the active DRL defender. Note that, for simplicity, 
in multi-agent red team settings, $c$ is deployed uniformly across all red agents such 
that they share identical behavioral logic throughout each episode.

\vspace{-2mm}
\subsection{Optimization via GRPO}
\vspace{-2mm}

To optimize the meta-agent $\pi_{\theta}$ under this formulation, we utilize Group Relative Policy Optimization (GRPO) \cite{preliminary1} as a surrogate objective for \autoref{eq:bandit}. The GRPO objective and the associated relative advantage are defined as:

\vspace{-3mm}
\begin{equation}
\label{eq:grpo}
\begin{aligned}
    \mathcal{J}_{\text{GRPO}}(\theta) &= \mathbb{E} \Bigg[ \frac{1}{G} \sum_{i=1}^{G} \min \Bigg( \frac{\pi_\theta(c_i | C)}{\pi_{\theta_{\text{old}}}(c_i | C)} A_i, \text{clip} \left( \frac{\pi_\theta(c_i | C)}{\pi_{\theta_{\text{old}}}(c_i | C)}, 1-\epsilon, 1+\epsilon \right) A_i \Bigg) \\
    &\quad - \beta \mathbb{D}_{\text{KL}} (\pi_\theta \| \pi_{\text{ref}}) \Bigg],  A_i = \frac{r_i - \text{mean}(r_1, \dots, r_G)}{\text{std}(r_1, \dots, r_G)}
\end{aligned}
\end{equation}
\vspace{-3mm}

For each context $C$, GRPO samples a group of red policies $\{c_1, \dots, c_G\}$ from the old policy $\pi_{\theta_{\text{old}}}$. The advantage $A_i$ is computed by normalizing the verifiable rewards $r_i$ relative to the group performance, where $\beta$ and $\epsilon$ are hyperparameters for stability, and $\pi_{\text{ref}}$ is a frozen reference model to ensure training stability. This objective allows the meta-agent to iteratively refine its strategic reasoning, discovering effective attack logics against adaptive DRL defenders through direct environmental feedback.

\vspace{-3mm}
\section{Trident}
\vspace{-3mm}
Optimizing $J(\theta)$ via GRPO requires a distribution of high-quality red-blue 
interaction trajectories to sample contexts $C$ from. However, no existing resource 
provides such trajectories against active DRL defenders. Trident addresses this 
by providing both the dataset and the benchmark infrastructure needed to realize 
RLVR training in this setting.

\vspace{-3mm}
\subsection{Dataset preparation}
\vspace{-3mm}

As summarized in \autoref{tab:benchmark_comparison}, existing LLM-driven 
cybersecurity benchmarks lack long-horizon adversarial dynamics. Large-scale 
datasets predominantly provide static security corpora~\citep{yu-etal-2025-primus} 
or QA pairs for supervised fine-tuning~\citep{suryanto2026redsage, yu-etal-2025-primus, 
ctibench, wan2024cyberseceval}, while interactive benchmarks focus primarily on 
isolated CTF~\citep{zhuo2026cyberzero, intercodectf} or single-target Common 
Vulnerabilities and Exposures (CVE) exploitation~\citep{zhu2025cvebench}. None of 
these capture the continuous, multi-step interactions required to defeat active 
network defenders. To fill this gap, we construct $\mathcal{D}$, a large-scale dataset of over 13,000 high-fidelity red-blue interaction trajectories spanning CAGE 4 and CyberWheel, whose construction is detailed below.

\begin{table}[htbp]
    \vspace{-3mm}
    \centering
    \caption{Comparison of \textbf{Trident} Dataset against existing benchmarks. By focusing on core capabilities, Trident uniquely provides a massive-scale dataset combined with an interactive environment supporting Agentic evaluation, RLVR, and Active DRL defenses.}
    \label{tab:benchmark_comparison}
    \resizebox{\columnwidth}{!}{
    \begin{tabular}{llcccc}
        \toprule
        \textbf{Benchmark} & \textbf{Task} & \textbf{Scale (Train / Total)} & \textbf{Agentic?} & \textbf{RLVR?} & \textbf{Active Defense?} \\
        \midrule
        NYU CTF \cite{shao2024nyuctfbench}           & CTF              & - / 200   & \cmark & \xmark & \xmark \\
        InterCode-CTF \cite{intercodectf}       & CTF              & - / 100         & \cmark & \xmark & \xmark \\
        Cybench \cite{zhang2025cybench}               & CTF              & - / 40          & \cmark & \xmark & \xmark \\
        Cyber-Zero \cite{zhuo2026cyberzero}  & CTF              & 6,188 / 6,188          & \cmark & \xmark & \xmark \\

        \midrule
        CVE-bench \cite{zhu2025cvebench}            & CVE Exploitation & - / 509         & \cmark & \xmark & \xmark \\

        CyberSecEval 3 \cite{wan2024cyberseceval} & Security Eval. & - / 6 & \cmark & \xmark & \xmark \\
        
        CTI-bench \cite{ctibench}            &  Security QA    & 2,500 / 2,500           & \xmark & \xmark & \xmark \\
        RedSage \cite{suryanto2026redsage}   & Security QA      & 266K / 266K     & \xmark & \xmark & \xmark \\

        PRIMUS \cite{yu-etal-2025-primus}           & Security Corpus      & 674K / 674K     & \xmark & \xmark & \xmark \\
        \midrule
        CyberBattleSim \cite{msft:cyberbattlesim} & Network Attack   & N/A           & \xmark & \xmark & \xmark \\

        CybORG CAGE 4 \cite{kiely2025exploring}                 & Network Defense  & N/A           & \xmark & \xmark & \cmark \\

        CyberWheel \cite{cyberwheel} & Network Attack/Defense   & N/A           & \xmark & \xmark & \cmark \\

        \midrule
        \rowcolor{gray!10}
        \textbf{Trident (Ours)}              & \textbf{Network Attack} & \textbf{13,056$+$ / 13,744$+$} & \cmark & \cmark & \cmark \\
        \bottomrule
    \end{tabular}
    }

\end{table}

\noindent \textbf{Sandbox infrastructure :}
To enable high-concurrency parallel rollouts, we decouple policy execution 
from the simulation runtime via a client-server architecture. Trident operates on the client side, transmitting generated Python policies 
to isolated, containerized sandbox servers that execute them natively against 
live DRL defenders and return verifiable execution logs with deterministic 
reward signals. Each sandbox exposes a standardized RESTful interface, 
allowing concurrent rollouts across independent simulator instances without 
shared state, guaranteeing reproducibility and integrity of $\mathcal{D}$.


\noindent \textbf{Trajectory rollout:}
We leverage a diverse suite of $N$ red agent against 
a fixed DRL blue defender (Details in \autoref{app_b:heuristic}), iteratively sampling a heuristic 
$c_j \sim \text{Uniform}(1, N)$ and a randomized seed $s_i \sim \mathcal{S}$ 
to generate trajectories, collected as:
\begin{equation}
    \mathcal{D} = \left\{ \tau^{\prime}_{i} = \left( o_t^{(i)}, a_t^{(i)} \right)_{t=0}^{T} \sim P(\cdot \mid c_j, s_i) : 1 \leq i \leq n, \hspace{2mm} j \sim \text{Uniform}(1, N) \right\}
\end{equation}


\begin{figure}[t!]
    \vspace{-5mm}
    \centering
    \begin{subfigure}[t]{0.51\linewidth}
        \centering
        
        \caption{CAGE4}
        \vspace{-1mm}
        \includegraphics[width=\linewidth]{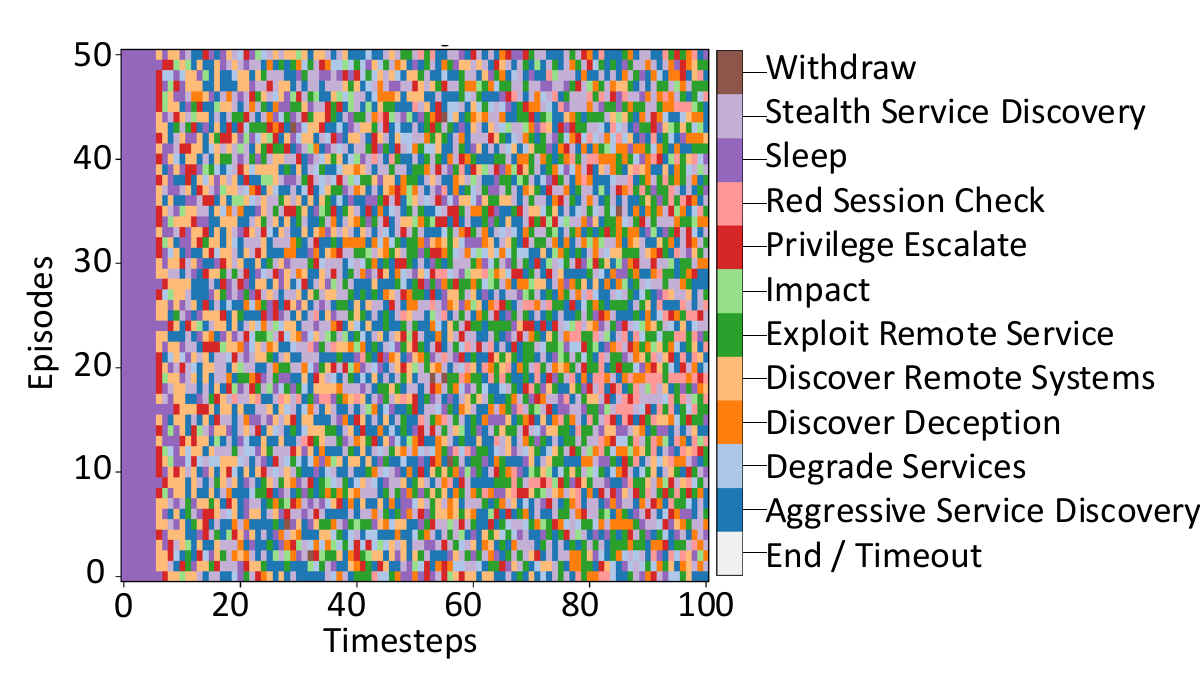}
        \label{fig:diversity_a}
    \end{subfigure}
    \hfill
    \begin{subfigure}[t]{0.48\linewidth}
        \centering
        \caption{CyberWheel}
        \vspace{-1mm}
        \includegraphics[width=\linewidth]{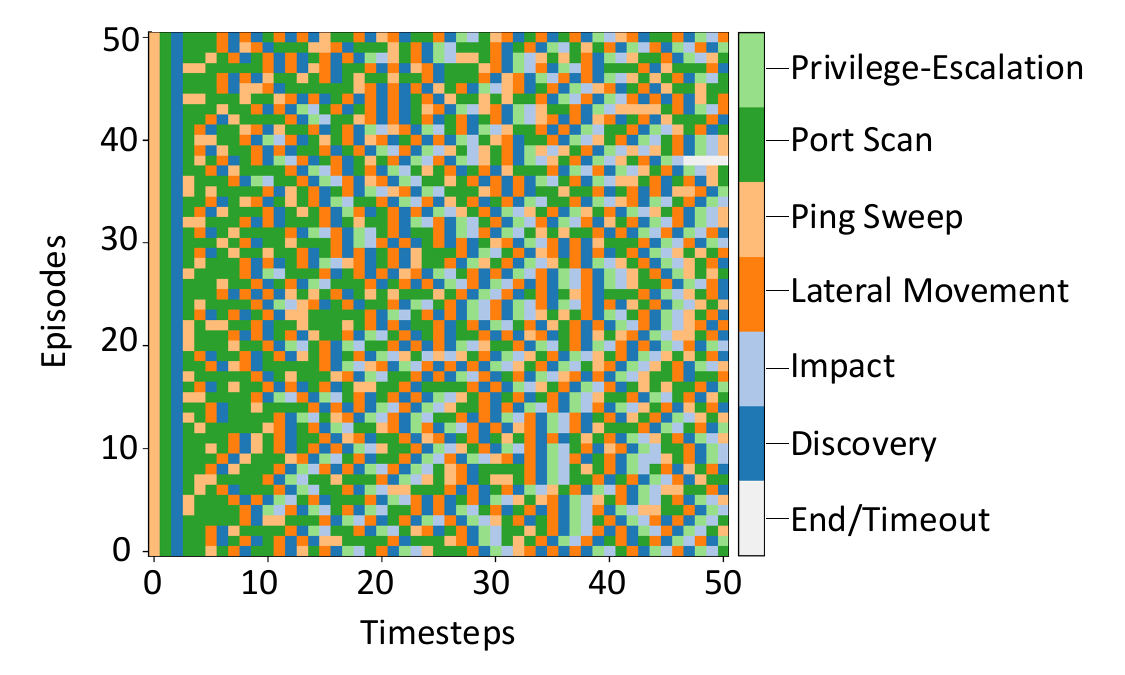}
        \label{fig:diversity_b}
    \end{subfigure}
    \vspace{-5mm}
    \caption{\textbf{Tactical branching in CybORG and CyberWheel.} Each row represents one of 50 sampled trajectories from a baseline heuristic against a fixed DRL defender. This seed variation ensures diverse data for RLVR training.}
    \label{fig:diversity}
    \vspace{-6mm}
\end{figure}

This rollout process yields 3,556 training and 188 validation trajectories for 
CyberWheel, and 9,500 training and 500 validation trajectories for CAGE 4, 
totaling 13,744 samples. As the framework supports arbitrary seed randomization, 
additional trajectories can be generated on demand. As illustrated in 
\autoref{fig:diversity}, even fixed heuristic logic produces significant tactical 
branching across seeds and defender interventions, capturing the complexity of the 
adversarial landscape.

\noindent \textbf{Natural language log serialization: }
Each trajectory $\tau^{\prime}_{i}$ is serialized into a structured natural language logs via a fixed, deterministic template. Each timestep encodes the agent's partial observation, differential host intelligence updates, and executed actions at every step $t$, followed by an episode-level impact summary, as illustrated in \autoref{fig:prompt_template}.
This prompt dataset bootstraps the meta-agent with a structural map of the adversarial landscape. A more detailed 
analysis of the Trident Dataset, including its coverage, diversity, outcome 
distribution, redundancy, and train--validation consistency, is provided 
in~\autoref{app:f_dataset}.

\vspace{-2mm}
\subsection{Trident Agentic}
\vspace{-2mm}
Trident Agentic decomposes the attack agent into three specialized modules, 
each with a distinct role: compressing observations, reasoning about 
strategy, and generating executable code. Following prior multi-agent 
designs~\citep{agent_design1, agent_design2, agent_design3, agent_design4}, 
this separation of concerns isolates the RL optimization signal to a single 
trainable strategic component, while keeping the remaining modules frozen.


\noindent \textbf{Log summarizer:} A hybrid module that uses a rule-based parser combined with a frozen LLM to compress a prohibitive amount of serialized logs (\autoref{fig:prompt_template}). To ensure the input remains within token limits without losing critical information, the parser identifies and preserves all ``milestone'' steps—such as successful exploits, privilege escalations, or service impacts—while randomly sampling from failed attempts and reconnaissance noise. This heuristic filtering ensures that the most informative parts of the attack trajectory are always presented to the LLM. The model then synthesizes these filtered logs into a dense semantic summary for the Planner.

\begin{wrapfigure}{r}{0.4\textwidth}
\centering
\includegraphics[width=\linewidth]{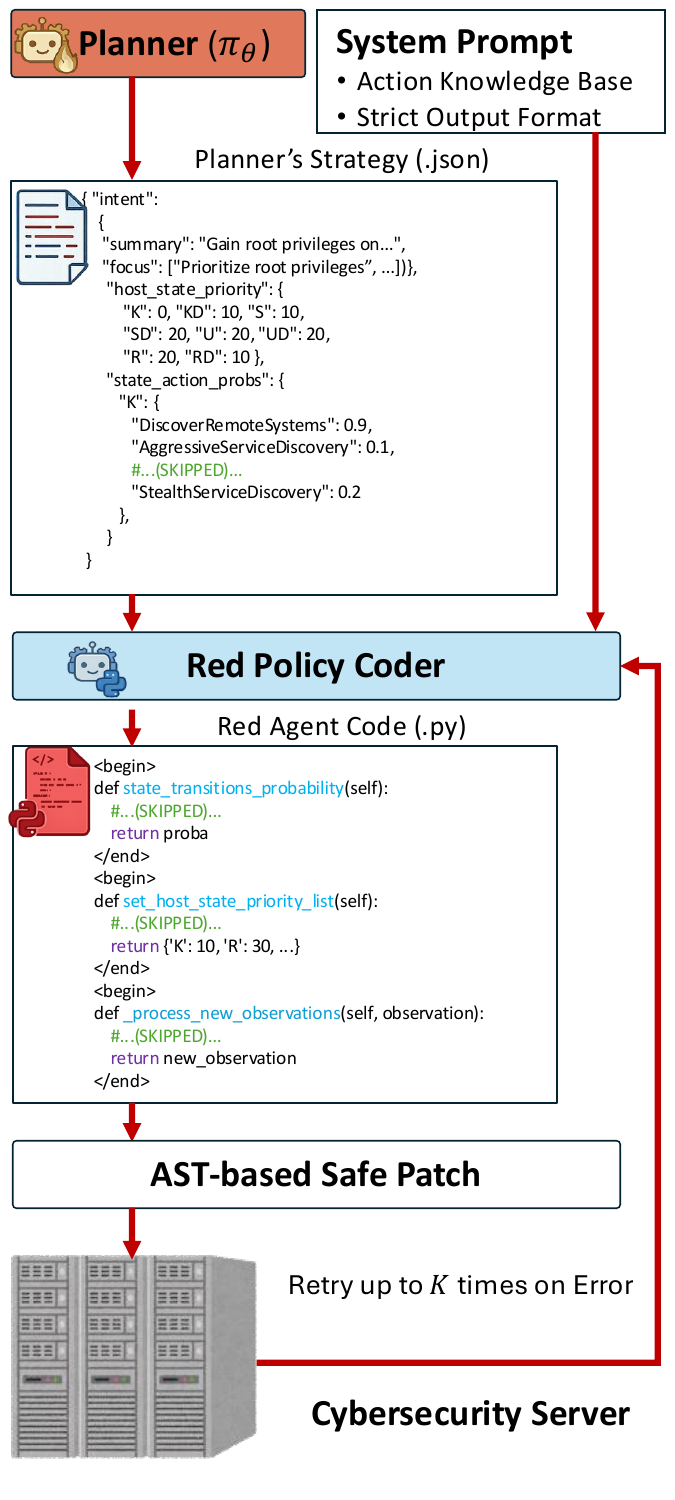}
\vspace{-6mm} 
\caption{Planner-Coder Pipeline.}\label{fig:prompt_design}
\vspace{-13mm} 
\end{wrapfigure}
\noindent \textbf{Planner} ($\pi_\theta$): The sole trainable component of our architecture. Conditioned on the condensed summary as its partially observable state, it formulates abstract tactical logic, outputting a structured JSON payload as its strategic action. By delegating exact programming syntax to the Coder, the Planner focuses exclusively on high-level decision-making, smoothing the optimization landscape. Its weights are updated directly via GRPO (\autoref{eq:grpo}) driven by verifiable execution outcomes, ensuring the policy learns generalized attack strategies strictly from observable feedback without any privileged environment access.

\noindent \textbf{Coder:} As illustrated in \autoref{fig:prompt_design}, a frozen model translates the Planner's strategic intent into executable Python code. 
Following recent advancements in robust LLM-driven code generation~\citep{ast1, ast2}, we incorporate an Abstract Syntax Tree (AST)-based patching mechanism to verify structural integrity and ensure execution stability before deployment. 

Furthermore, an iterative self-repair loop is employed to mitigate minor syntax errors that could disrupt the learning. A sandbox performs a fast-pass sanity check; if an execution error occurs, the traceback is provided to the Coder for correction. 
Permitting up to $K$(=3) repair attempts absorbs trivial syntax errors, ensuring the Planner is not penalized for incidental implementation noise. However, if the \textit{Coder} exhausts all retries, this indicates the Planner's strategy is fundamentally unimplementable, and the resulting execution penalty is intentionally applied.

\vspace{-2mm}
\subsection{Domain-grounded prompt design}
\vspace{-2mm}

A critical challenge in applying LLMs to cybersecurity is preventing hallucinated actions and ensuring the agent operates strictly within the observable state. To address this, each module in Trident is governed by heavily engineered. 
First, we inject an action knowledge base directly into the system prompts, providing the LLMs with explicit prerequisites and expected outcomes for complex killchain actions. Second, we enforce Strict API Boundaries. The Coder and Planner are explicitly forbidden from accessing unobservable simulation internals (e.g., querying decoy flags). They are constrained to use only legal APIs, forcing them to rely on behavioral inference rather than omniscient state access. 
Finally, to guarantee deterministic pipeline execution, we utilize rigid output schemas—forcing the Planner to generate abstract strategy JSON, and constraining the Coder to output exactly delineated Python blocks.






\vspace{-2mm}
\subsection{Reward design}
\vspace{-2mm}

To accurately evaluate the generated algorithmic policy $c$, we design a dense, verifiable reward 
function derived directly from the execution logs of each trajectory $\tau$. Let $\phi(e, o_{0:T})$ 
count occurrences of event $e$ across the observation sequence of length $T$, and let $\mathcal{E}^{+}$, 
$\mathcal{E}^{-}$ denote positive milestone events (e.g., system impacts, privilege escalations) and 
negative events (e.g., failed exploits, invalid actions), respectively. We define the tactical reward 
components as:
\begin{equation}
    R^{+} = \sum_{e \in \mathcal{E}^{+}} q_e \phi(e, o_{0:T}) 
    + \frac{r}{T}\sum_{e \in \mathcal{E}^{+}} \phi(e, o_{0:T}), 
    \qquad 
    R^{-} = \sum_{e \in \mathcal{E}^{-}} s_e \phi(e, o_{0:T})
\end{equation}
where $q_e$ is a per-event weight reflecting the relative importance of each positive milestone, 
$r$ is a scalar efficiency coefficient that normalizes total achievements by episode length $T$, 
and $s_e$ is a per-event penalty weight controlling the severity of each negative event. 
Thus $R^{+}$ accumulates weighted killchain milestones augmented by an efficiency bonus, 
and $R^{-}$ penalizes wasteful behavior.

Crucially, to penalize syntactically invalid or non-executable logic, the final episodic reward is 
determined via a conditional branch. If the generated policy $c$ fails to execute due to unrecoverable 
syntax or runtime errors, the episode is aborted and a strict execution penalty $p$ is applied. 
Otherwise, the policy receives a tactical reward clipped within $[R_{\min}, R_{\max}]$:
\begin{equation}
    \label{eq:reward}
    R(c, o_{0:T}) = 
    \begin{cases} 
        p & \text{if } c \text{ fails to execute or violates constraints} \\
        \text{clip}\!\left(R^{+} - R^{-},\ R_{\min},\ R_{\max}\right) & \text{otherwise}
    \end{cases}
\end{equation}
Full specifications of $\mathcal{E}^{+}$, $\mathcal{E}^{-}$, and $p$ 
for each environment are provided in \autoref{app:cage4_reward}, \autoref{app:cyberwheel_reward}, and \autoref{app:penalties}, respectively.

\section{Experiments}
\label{sec:exp}
\begin{table}[h]
\centering
\caption{Performance comparison across CAGE4 and CyberWheel settings.
Lower returns indicate worse blue-agent performance. Values are reported
as mean $\pm$ standard deviation. Parenthesized values denote the relative
degradation ($\Delta$) from the Default Agent; positive values indicate
stronger attacks. The final column averages $\Delta$ over the four CAGE4
defenders only.}
\label{tab:cage4_cyberwheel}
\resizebox{\linewidth}{!}{
\begin{tabular}{lcccccc}
\toprule
\multirow{2}{*}{\textbf{Method}}
& \multicolumn{4}{c}{\textbf{CAGE4}}
& \multicolumn{1}{c}{\textbf{CyberWheel}}
& \multirow{2}{*}{\shortstack{\textbf{Average CAGE4}\\\textbf{Degradation ($\Delta$)}}} \\
\cmidrule(lr){2-6}
& \textbf{GNN~\cite{cybermonic}}
& \textbf{HMARL3~\cite{hmarl}}
& \textbf{HMARL4~\cite{hmarl}}
& \textbf{MARL~\cite{victim_marl}}
& \textbf{PPO (NN)~\cite{cyberwheel}}
& \\
\midrule

Default Agent
& -202.34 $\pm$ 1.42
& -132.63 $\pm$ 0.87
& -940.87 $\pm$ 15.75
& -171.09 $\pm$ 4.32
& 4574.59 $\pm$ 3.65
& -- \\

GPT-4o~\cite{hurst2024gpt}
& -351.43 $\pm$ 65.30 \gain{(73.68\%)}
& -191.56 $\pm$ 42.17 \gain{(44.44\%)}
& -844.99 $\pm$ 196.51 \loss{(-10.19\%)}
& -323.91 $\pm$ 9.85 \gain{(89.33\%)}
& 1919.31 $\pm$ 149.59 \gain{(58.04\%)}
& \gain{49.32\%} \\

Zero-shot
& -498.74 $\pm$ 65.85 \gain{(146.48\%)}
& -598.09 $\pm$ 26.15 \gain{(350.96\%)}
& -1701.01 $\pm$ 113.20 \gain{(80.79\%)}
& -435.84 $\pm$ 29.82 \gain{(154.75\%)}
& 2107.84 $\pm$ 59.64 \gain{(53.92\%)}
& \gain{183.25\%} \\

CoT
& -367.19 $\pm$ 19.58 \gain{(81.47\%)}
& -200.09 $\pm$ 22.76 \gain{(50.87\%)}
& -737.94 $\pm$ 171.87 \loss{(-21.57\%)}
& -340.55 $\pm$ 42.79 \gain{(99.05\%)}
& 1729.16 $\pm$ 69.83 \gain{(62.20\%)}
& \gain{52.46\%} \\

DRL Red
& -311.11 $\pm$ 25.24 \gain{(53.76\%)}
& -271.88 $\pm$ 9.47 \gain{(104.99\%)}
& -314.52 $\pm$ 14.99 \loss{(-66.57\%)}${}^{\dagger}$
& -262.13 $\pm$ 13.27 \gain{(53.22\%)}
& \textbf{-66.35 $\pm$ 27.52 \gain{(101.45\%)}}
& \gain{36.35\%} \\

\textbf{Trident Agentic (Ours)}
& \textbf{-2083.68 $\pm$ 110.89 \gain{(929.79\%)}}
& \textbf{-849.85 $\pm$ 6.46 \gain{(540.78\%)}}
& \textbf{-2241.72 $\pm$ 194.82 \gain{(138.26\%)}}
& \textbf{-1716.73 $\pm$ 33.28 \gain{(903.43\%)}}
& \textbf{35.70 $\pm$ 48.89 \gain{(99.22\%)}}
& \textbf{\gain{628.07\%}} \\

\bottomrule
\end{tabular}
}
\vspace{-2mm}
\end{table}

To evaluate the effectiveness and robustness of Trident Agentic, we conduct our experiments within multiple environments with different victim models. We design our evaluation to answer two primary Research Questions (RQs):
(\textbf{RQ1}) Can the algorithmic policy generated by Trident Agentic defeat state-of-the-art DRL blue agents where traditional heuristic attackers fail?
(\textbf{RQ2}) Is RLVR training beneficial for generating effective attacks against DRL blue agents?

\subsection{Experimental setup and baselines}

\noindent \textbf{Implementation Details:}
We used Qwen2.5-7B-Instruct for our planner policy($\pi_\theta$), Qwen2.5-14B-Instruct-AWQ~\cite{qwen2, qwen2.5} for summarizer and Qwen2.5-Coder-7B-Instruct~\cite{hui2024qwen2} for coder agent. 
We used a single NVIDIA-H200 GPU to host frozen LLM agents and 2 NVIDIA-H200 GPUs for training our planner policy. 
The group size of GRPO ($G$) is set to 8, learning rate of $5.0\times 10^{-7}$ and batch size of 2. The KL penalty term ($\beta$) is set to $0.05$ for CyberWheel and $5.0\times 10^{-4}$ for CAGE4, respectively. The implementation was mainly built upon verl~\cite{verl} for the RLVR training loop, utilizing vllm~\cite{vllm} for efficient policy generation and Ray~\cite{ray} for distributed orchestration.
\begin{wrapfigure}{r}{0.5\textwidth}
\vspace{-1.2em}
\centering

\begin{tcolorbox}[
    width=\linewidth,
    colback=green!2!white,
    colframe=green!50!black,
    title=\textbf{CAGE4: Dynamic State Priority Adaptation},
    boxsep=2pt,
    left=4pt,
    right=4pt,
    arc=3pt
]

{\small \textbf{[Default Code (CAGE4)]}}\\[-2pt]
{\scriptsize \ttfamily
def set\_host\_state\_priority\_list(self):\\
\hspace*{2mm}return None {\color{blue!70!black}{\# equal priority (default agent)}
}}

\vspace{3pt}
{\small \textbf{[Planner's Thought]}}\\[-2pt]
{ \scriptsize
Shift from discovery to exploitation. Prioritize high-value targets and
increase focus on privilege escalation and root compromise.
}

\vspace{3pt}
{\small \textbf{[Coder's Implementation]}}\\[-2pt]
{\scriptsize \ttfamily
def set\_host\_state\_priority\_list(self):\\
\hspace*{2mm}return \{\\
\hspace*{4mm}'K': 10, 'KD': 10, 'S': 10, 'SD': 10,\\
\hspace*{4mm}'U': 20, 'UD': 20, 'R': 30, 'RD': 30\\
\hspace*{2mm}\}
}
\end{tcolorbox}
\vspace{-4mm}
\caption{CAGE4 policy generation example. Trident Agentic coder introduces a structured priority that shifts focus toward high-value targets.}
\vspace{-4mm}

\label{fig:cage4_policy}
\end{wrapfigure}
\noindent \textbf{Environment Setup:} Both CAGE 4~\cite{kiely2025exploring} and CyberWheel~\cite{cyberwheel} model enterprise networks as adversarial POMDPs where a killchain-based red agent attacks hosts across subnets while a DRL blue agent defends in real-time — producing deterministic, verifiable reward signals from discrete-event logs. Despite this shared structure, they differ in network topology, defender architecture, and mission complexity. CAGE 4 features a multi-agent military network with dynamic mission phases and cooperative blue defenders, while CyberWheel grounds its red agent actions in real-world MITRE ATT\&CK~\cite{mitreMITREATTampCKxAE} techniques, providing higher attack realism. We evaluate across both to test the generalizability of Trident. Full environment details are provided in \autoref{app:a_environ}.

\noindent \textbf{Victim Models:} We evaluate the performance of Trident Agentic against three state-of-the-art DRL blue agents.
``GNN''~\cite{cybermonic, kiely2025exploring} represents the network as a temporal attributed
graph processed by a graph convolutional network optimized by Proximal Policy Optimization (PPO)~\cite{ppo},
with actions formulated as graph edits on nodes and edges, enabling zero-shot generalization to unseen topologies.
``H-MARL''~\cite{hmarl} employs a hierarchical architecture optimized by PPO that decomposes cyber defense
into sub-tasks, coordinated by either an expert-rule or a learned master policy across five blue agents.
``MARL''~\cite{victim_marl} is a decentralized multi-agent PPO baseline where each blue agent maintains
a separate actor-critic trained solely on its local observations.
All three are evaluated in CAGE 4. For CyberWheel, we evaluate against ``PPO(NN)''~\cite{cyberwheel},
a PPO-based blue agent from the official CyberWheel repository trained to minimize server downtime against a heuristic red agent.


\noindent \textbf{Dataset and Training Split:} For each target DRL blue agent, we trained a dedicated red policy ($\pi_\theta$) via RLVR exclusively on the training split of the Trident Dataset. Specifically, policies evaluated in CAGE 4 are trained solely on the CAGE 4 training split, and those evaluated in CyberWheel are trained solely on the CyberWheel training split. The performance reported in our evaluation is then derived from the corresponding validation splits to ensure a rigorous assessment of generalization across unseen network states and defensive configurations.

\noindent \textbf{Baselines:}
We compare Trident against a rule-based default agent, three alternative
planner configurations that isolate the contribution of RLVR training, and
a simulator-native DRL baseline that tests whether generic policy adaptation
alone is sufficient. The ``Default Agent'' uses the default
\texttt{FiniteStateRedAgent} from the CybORG CAGE 4 repository for CAGE 4
and the \texttt{BFSServerDowntime} strategy for CyberWheel. The ``DRL Red
(PPO)'' baseline trains a separate PPO red agent against each frozen blue
defender using the simulator-native observation and action interfaces. Unlike
Trident, it directly learns a step-wise numerical policy without the Log
Summarizer, Planner, Coder, or GRPO components. This baseline therefore
distinguishes the benefit of Trident's agentic policy-generation framework
from generic adaptation to a fixed defender. Its architecture, training
budget, and evaluation details are provided in \autoref{app:ppo_baseline}.

\begin{wrapfigure}{r}{0.56\textwidth}
\centering
\vspace{-5mm}
\begin{tcolorbox}[
    width=\linewidth,
    colback=blue!2!white,
    colframe=blue!70!black,
    title=\textbf{CyberWheel: Decoy Avoidance},
    boxsep=1pt,
    left=3pt,
    right=3pt,
    arc=3pt
]
{\small \textbf{[Default Red Agent]} \\
Simple BFS prioritizing unimpacted servers.}\\
{\small \textbf{[Planner's Thought]}}\\
{\scriptsize
``Based on the analysis, we can summarize the key issues with the current
strategy:
1. Fixation on single hosts, leading to redundant actions.
2. Inefficient exploitation of the early window.
3. Lack of effective behavioral detection for decoys.
We need to reformulate our strategy to ensure that the agent can efficiently
reach and impact unimpacted servers within the early window and avoid fixating
on already impacted or suspicious hosts. Additionally, we should develop a
mechanism to detect and avoid decoys.''}\\
{\small \textbf{[Generated Logic (Python)]}}\\
{\scriptsize \ttfamily \color{black!85}
class ImprovedRedStrategy:\\
\hspace*{2mm}{\color{blue!70!black}\# Learned parameter avoiding decoys}\\
\hspace*{2mm}DECOY\_THRESHOLD = 3 \\
\hspace*{2mm}def select\_target(cls, agent):\\
\hspace*{4mm}if history.svc\_count == 0:\\
\hspace*{6mm}cls.failures[host] += 1\\
\hspace*{6mm}if cls.failures >= cls.THRESHOLD:\\
\hspace*{8mm}return cls.select\_next\_target()
{\color{red!70!black}\# Abort} \\
\hspace*{2mm}{\color{red!90!black}----------Skipped----------}
}
\end{tcolorbox}
\vspace{-3mm}
\caption{Example generated logic for decoy avoidance in CyberWheel that
abandons targets.}
\vspace{-8mm}
\label{fig:wheel_sample}
\end{wrapfigure}

The remaining LLM-based configurations vary only the planner. ``Zero-shot''
uses the same base model without task-specific training; ``CoT''
\cite{kojima2022large} uses an identical base model and prompt augmented with
a chain-of-thought instruction; and ``GPT-4o'' uses GPT-4o
\cite{hurst2024gpt} to evaluate the effect of frontier-model capacity without
task-specific training. These LLM-based configurations use identical
observations and evaluation protocols, and all prompts are identical except
for the additional CoT instruction. The PPO baseline uses its native numerical
interface but is evaluated using the same benchmark-specific protocol.

\noindent \textbf{Evaluation Metrics:} We adopt the blue agent's cumulative episodic reward as our primary metric, where a lower value indicates more effective red agent attacks, consistent with prior DRL defense benchmarks. This captures disruption to green agent activity across mission phases in CAGE 4, and defender success via decoys in CyberWheel. CAGE 4 results are reported at training step 150, averaged over 500 randomized episodes of 500 steps each ($T$=500 in \autoref{eq:bandit}); CyberWheel results are reported at training step 1,000, averaged over 188 randomized episodes of 100 steps each. All baselines are reported as the mean and standard deviation across 3 independent runs. We note a deliberate asymmetry favoring baselines: GPT-4o, Zero-shot, and CoT metrics are computed \textit{exclusively over successful executions}, filtering out fatal errors, while Trident's evaluation is strictly comprehensive, absorbing all execution failures and their associated penalties.

\subsection{Results}
\noindent \textbf{Quantitative Analysis: } Table \ref{tab:cage4_cyberwheel} presents cumulative rewards against all defensive baselines. Heuristic red agents are weak evaluators: blue agents trained exclusively against them condition on their fixed behavioral signatures, leaving systematic blind spots invisible under current evaluation practices. Trident exploits these blind spots by dynamically synthesizing executable attack scripts tailored to the evolving network state, degrading the four CAGE 4 defenders by an average of 628.07\% and eliminating over 99\% of the CyberWheel defender's reward, affirmatively answering \textbf{RQ1}. Notably, the Zero-shot SLM outperforms GPT-4o in several settings, which we attribute to prompt optimization bias: our scaffolding was iteratively developed around the 7B model, and correcting this disadvantage via extensive prompt-tuning was computationally infeasible over thousands of interactive DRL episodes. Despite this inadvertent advantage to the Zero-shot baseline, Trident Agentic still substantially surpasses it: static prompt engineering alone only partially compromises DRL defenders, while RLVR training provides a decisive advantage, answering \textbf{RQ2}.

\noindent \textbf{Qualitative Analysis:}
\autoref{fig:cage4_policy} and \autoref{fig:wheel_sample} illustrate 
emergent attack behaviors autonomously discovered by Trident Agentic 
through RLVR. In CyberWheel, Trident Agentic identifies the baseline agent's naive BFS strategy as vulnerable to honeypots and autonomously synthesizes a decoy-avoidance mechanism that aborts attacks on suspicious hosts after repeated failures. In CAGE 4, rather than treating all network states equally, Trident Agentic learns to prioritize high-value states such as Root and User access over basic reconnaissance states. These behaviors confirm that the ``Code-as-Policies'' paradigm provides the expressivity necessary to discover novel, environment-specific attack 
vectors that static heuristics cannot represent.

\noindent \textbf{Ablation Studies:} Due to space constraints, detailed ablation studies regarding the reward formulation and the design choices of our agentic components are deferred to \autoref{app:c_reward_ablation}.






\section{Limitations}

While Trident Agentic demonstrates exceptional efficacy against state-of-the-art DRL defenses, two engineering choices limit out-of-the-box generalization. First, the framework depends heavily on domain-specific prompt engineering: the tripartite architecture relies on hand-crafted templates and rule-based log parsers, meaning adaptation to new simulation engines requires manual recalibration of these scaffolds. Second, the reward design remains environment-dependent; although the underlying concepts are universal, their implementation is tied to the specific log formats and action spaces of CAGE 4 and CyberWheel. Future work will focus on automating log parsing and reward discovery to move toward a truly zero-shot, adaptive red agent.

\section{Broader impacts and ethical considerations}
\label{sec:ethical}
This work demonstrates that sophisticated offensive capabilities against state-of-the-art DRL defenses can be realized with compact, open-weight SLMs, introducing a dual-use risk that will grow as simulators more closely approximate real-world networks. Assuming adversaries will not independently develop these methods is not a viable security posture. We argue that transparency is preferable to obscurity: by openly demonstrating Trident Agentic's capabilities, we expose a critical evaluation gap and call on the community to develop equally adaptive, RL-driven blue agents. This framework provides the automated adversaries and verifiable metrics needed to train more robust cyber defenses before these attack strategies are independently deployed.

\section{Conclusions}
In this paper, we introduced \textbf{Trident}, a dynamic red teaming benchmark comprising isolated sandbox servers, red-blue interaction datasets, and \textbf{Trident Agentic}, a tripartite RLVR agentic architecture to reframe autonomous red teaming as a contextual bandit problem. Our evaluations reveal a consistent brittleness in existing autonomous defenders: a single trainable 7B LLM planner achieves an average \textbf{628\%} reduction in blue-agent performance in CAGE 4 and erases over \textbf{99\%} of defender reward in CyberWheel, while autonomously discovering novel behaviors such as decoy avoidance and adaptive state prioritization. Beyond its immediate contributions, Trident opens several exciting future directions: its sandbox infrastructure naturally supports a self-improving curriculum where trained red agent policies generate richer adversarial trajectories to bootstrap progressively stronger attackers, whose pressure drives more robust blue agents in a game-theoretic co-evolution loop, while its RESTful design invites integration with higher-fidelity environments toward real-world autonomous cyber defense.

\bibliography{iclr2027_conference}
\bibliographystyle{iclr2027_conference}

\appendix
\section*{Appendix}
\renewcommand{\sectionautorefname}{Appendix}
\renewcommand{\subsectionautorefname}{Appendix}%
\renewcommand{\subsubsectionautorefname}{Appendix}%

\section{Related works}
\subsection{Cybersecurity and LLMs}

LLM-based cybersecurity research has evolved from static evaluations, including vulnerability detection~\citep{devign, zhu2025cvebench}, threat intelligence QA~\citep{ctibench}, and secure code evaluation~\citep{purplellama, wan2024cyberseceval, bhatt2024cyberseceval, guo2024redcode}, toward interactive agentic systems~\citep{llm_hacker1, llm_hacker2, llm_hacker3, llm_hacker4, llm_hacker5}, driven by competitions such as DARPA AIxCC~\citep{aicyberchallengeOverviewx2013, liu2024cyberbench} and evaluated on CTF challenges and sandboxes~\citep{abramovich2025enigma, zhang2025cybench, shao2024nyuctfbench, iclr26_cybergym, zhang2025defenderbench}. However, these frameworks feature static targets without active defenders, while traditional DRL simulators~\citep{kiely2025exploring, msft:cyberbattlesim} operate on numeric vector spaces unsuitable for LLM-native reasoning, and recent cybersecurity LLMs~\citep{zhuo2026cyberzero, yu-etal-2025-primus, suryanto2026redsage} rely on offline Supervised Fine Tuning (SFT) datasets that cannot support online RL against adaptive defenders. Trident bridges this gap by providing both an interactive text-native environment and a large-scale hybrid dataset for RLVR training, a need made urgent by the autonomous zero-day exploitation capabilities now demonstrated by frontier models such as Claude Mythos~\citep{anthropicClaudeMythos}. The closest concurrent work to ours is~\citep{trident_pakuri}, which also uses GRPO for adversarial red teaming. However, Trident trains executable code-as-policy agents against active DRL defenders in network-security environments with environment-verifiable rewards, while~\citet{trident_pakuri} focuses on prompt-based jailbreaks against LLMs.

\vspace{-3mm}
\subsection{Code-as-Policy and RLVR in agentic LLMs}
\vspace{-3mm}

Modern agentic frameworks formulate complex tasks as POMDPs using step-by-step prompting strategies in ReAct~\citep{yao2022react}-based manner~\citep{agent_design3, agent_design4}, where an agent interleaves reasoning traces with discrete tool calls to iteratively progress toward a goal. However, this approach is bottlenecked by the engineering overhead of designing universal tool wrappers across heterogeneous simulators and is vulnerable to error compounding in cybersecurity. The ``Code-as-Policies'' paradigm~\citep{code_as_policies, wei2025swerl} overcomes these limitations by generating executable programs that embed complex control flows directly into actions, eliminating brittle tool APIs. Recent works have integrated LLMs into RL pipelines for agentic reward design~\citep{ma2024eureka} and \cite{jiang2025agentic} adapted this into cyber defense environments, but these deploy LLMs as auxiliary modules rather than as primary policy generators. In contrast, Trident Agentic adopts the ``Code-as-Policies'' paradigm with iterative self-repair~\citep{ast1, ast2} to generate complete adversarial strategies directly. While RLVR has driven advancements in mathematics~\citep{preliminary1, datagen_evidence3} and competitive programming~\citep{zeng-etal-2025-acecoder_agent_example2}, Trident is, to the best of our knowledge, the first to successfully adapt it for training ``Code-as-Policies'' adversarial agents in cybersecurity.
\section{Environment Setup \& Reward Design}
\label{app:a_environ}
This section provides a comprehensive breakdown of the experimental testbeds, victim models, and reward formulations used to train and evaluate Trident Agentic. We deploy Trident Agentic across two distinct cybersecurity simulators: CybORG CAGE 4 (\autoref{app:cage4_setup}) and CyberWheel (\autoref{app:cyberwheel_setup}). For each environment, we detail the underlying network topology, the architecture of the victim DRL defenders (blue agents), and the baseline heuristic attackers (red agents). Crucially, we explain the exact mechanism by which Trident is integrated into these native simulators to dynamically generate and deploy its code-based adversarial policies (highlighted in \textcolor{blue}{blue} for clarity). Finally, we define the RLVR signals, outlining both the environment-specific tactical reward parsers and the universal system-level penalties (\autoref{app:penalties}) used to optimize Trident Agentic.
\subsection{CAGE 4}
\label{app:cage4_setup}

For our experiments utilizing the CybORG CAGE Challenge 4~\citep{kiely2025exploring} (CAGE 4) environment, we adopted its official enterprise network scenario, which is designed to emulate a complex, multi-subnet military network. While CyberWheel provides dynamically generated network graphs, CAGE 4 offers a complementary, rigorous testbed by maintaining a structurally fixed macro-topology while introducing severe micro-level variability. By randomizing the number of hosts, active services, and initial vulnerabilities in every episode, CAGE 4 allows us to evaluate the generated red policies for both strategic consistency across a stable network layout and tactical adaptability against shifting host-level configurations.

\subsubsection{Network topology}
The CAGE 4 environment simulates a highly segmented military enterprise architecture. The network is divided into four isolated enclaves connected via an internet node: two Deployed Networks (Network A and B), a Headquarters (HQ) Network, and an undefended Contractor Network that serves as the red agent's initial breach point. The topology enforces strict routing bottlenecks. For example, each Deployed Network contains a Restricted Zone facing the internet and an air-gapped Operational Zone accessible only through the Restricted Zone's router. Similarly, the HQ Network exposes a Public Access Zone, which internally routes to isolated Admin and Office networks.

To prevent deterministic policy memorization, the environment dynamically randomizes both the network configuration and the attack surface at the start of each episode. Each subnet is procedurally generated with a randomized allocation of 3 to 10 user workstations and 1 to 6 server hosts. Furthermore, while all hosts run baseline services like SSH, the environment probabilistically assigns supplementary services and injects exploitable vulnerabilities. This combination of dynamic node generation, strict hierarchical routing, and stochastic vulnerability distribution ensures that the Trident-generated red policy cannot rely on static killchains. Instead, it demands generalized reconnaissance, adaptive lateral movement, and conditional exploit logic.

\subsubsection{Agent configuration and Trident integration}
\textbf{Victim defenders (blue agents):} We utilized the official pre-trained weights for the HMARL and standard MARL defensive agents~\citep{victim_marl} available in the official GitHub repository of ~\citep{hmarl}. Additionally, we retrained the Graph Neural Network(GNN) defender using its official repository code~\citep{cybermonic}. This agent represents the network as a temporal attributed graph, processing partial observations through a GNN to output independent action probabilities for each of the five active blue agents. Following the official implementation, the policy was trained via PPO for 500,000 episodes (500 steps per episode) using a batch size of 2,500. The actor and critic networks utilize a 256-unit hidden layer and a 128-dimensional node embedding representation. Optimization was performed using learning rates of $3.0 \times 10^{-4}$ for the actor and $1.0 \times 10^{-3}$ for the critic, with a surrogate clipping coefficient of $0.2$ over 4 update epochs. All baseline blue policies remain strictly frozen during Trident's training and evaluation.

\textbf{Baseline attacker (red agent):} The default attacker in CAGE4 is the \texttt{FiniteStateRedAgent}. It employs an internal Finite State Machine (FSM) to orchestrate its attack killchain. The agent tracks each discovered host across 9 progressive compromise states (e.g., Known, User Shell, Root Shell). During execution, it selects a target host based on heuristic prioritization (e.g., defaulting to a 75\% bias toward servers) and selects its next action (e.g., \texttt{ExploitRemoteService} or \texttt{Impact}) by sampling from a hardcoded state transition probability matrix associated with the target's current state.

\textcolor{blue}{\textbf{Trident integration:} Rather than replacing the red agent entirely, Trident patches \texttt{FiniteStateRedAgent} by overriding three core methods: \texttt{state\_transitions\_probability}, \texttt{set\_host\_state\_priority\_list}, and \texttt{\_process\_new\_observations}. This narrows the generation space to high-level tactical decisions while delegating low-level killchain execution to the underlying framework, reducing syntax errors.}
\subsubsection{RLVR reward design and specifications}
\label{app:cage4_reward}
Consistent with our Code-as-Policy framework, we utilized a tactical log-based reward parser to evaluate Trident's performance in CAGE 4. As defined in \autoref{eq:reward}, the episodic reward $R(c, o_{0:T})$ is determined by a conditional branch. If the sandbox sanity check fails or the code triggers unrecoverable runtime errors during the self-repair loop, the episode is immediately terminated with a severe execution penalty ($p_{\text{exec}} = -5.0$, and $p_{\text{empty}} = -10.0$ for empty generation). 

For executable policies, the tactical reward is parsed directly from terminal logs:
\begin{itemize}
    \item \textbf{Positive Milestones ($\mathcal{E}^{+}$):} Rewards scale based on the absolute count of successful actions, breaking the traditional binary success metric. We assign weights of $q_{\text{impact}} = 0.2$, $q_{\text{root}} = 0.3$, $q_{\text{user}} = 0.1$, and $q_{\text{scan}} = 0.05$. To prioritize rapid exploitation, an efficiency multiplier $r = 2.0$ scales the reward based on the achievement-to-step ratio.
    \item \textbf{Negative Events ($\mathcal{E}^{-}$):} To penalize operational noise, the parser deducts points for invalid actions ($s_{\text{invalid}} = 0.05$) and failed exploits ($s_{\text{exploit\_fail}} = 0.1$).
\end{itemize}

The tactical portion of the reward is clipped within $[R_{\min} = -1.0, R_{\max} = 100.0]$. This separation of syntactic penalties and bounded tactical scores ensures that the GRPO gradient cleanly distinguishes between invalid code and poor strategic planning.

\subsection{CyberWheel}
\label{app:cyberwheel_setup}

For our CyberWheel~\citep{cyberwheel} experiments, we evaluate Trident on a 
15-host enterprise network topology. Although CyberWheel natively supports 
dynamically generated networks of varying scales (capable of simulating 
over 100,000 hosts) our primary objective was to validate Trident Agentic's 
generalizability beyond the CAGE 4 environment rather than exhaustively 
evaluate across topology sizes. Investigating policy performance across 
diverse network scales remains an compelling direction for future work.

\subsubsection{Network topology}
The 15-host network is centrally orchestrated around a single \texttt{core\_router} and partitioned into three functional subnets to replicate a standard enterprise architecture:
\begin{itemize}
    \item \textbf{DMZ Subnet :} Contains 5 outward-facing workstations (\texttt{dmz0} to \texttt{dmz4}). This serves as the primary entry point and decoy deployment zone for the blue agent.
    \item \textbf{User Subnet :} Contains 5 internal workstations (\texttt{host0} to \texttt{host4}).
    \item \textbf{Server Subnet :} Contains 5 critical business servers, specifically defining two proxy servers, one mail server, and two SSH jump servers (\texttt{server0} to \texttt{server4}).
\end{itemize}
To support complex lateral movement, explicit cross-subnet interfaces are configured, such as bridging \texttt{dmz1} directly to \texttt{server2}, and \texttt{host4} to \texttt{dmz2}.

\subsubsection{Agent configuration and Trident integration}
\label{app:cyberwhee_trident_integration}

\textbf{Victim Defender (blue agent):} To establish a rigorous victim model, we utilized a Proximal Policy Optimization (PPO) blue agent following the official CyberWheel implementation. Its policy employs a standard Actor-Critic architecture consisting of Multi-Layer Perceptrons (MLPs) with two hidden layers of 64 units each and ReLU activations. We maintained the official hyperparameters: a discount factor $\gamma = 0.995$, GAE $\lambda = 0.95$, and learning rates of $2.5 \times 10^{-4}$ with cosine restarts (minimum floor $1 \times 10^{-5}$). The surrogate objective was clipped at $0.3$, with an entropy coefficient of $0.02$, value function coefficient of $0.5$, and maximum gradient norm of $0.5$. This blue policy remains strictly frozen during Trident's training and evaluation.

\textbf{Baseline Attacker (red agent):} The default \texttt{ARTAgent}~\citep{cyberwheel} executes a hardcoded, deterministic killchain sequence (e.g., PingSweep $\rightarrow$ PortScan $\rightarrow$ Lateral Movement $\rightarrow$ Discovery $\rightarrow$ Privilege Escalation $\rightarrow$ Impact) grounded in MITRE ATT\&CK techniques. Crucially, its target selection logic is modularized into a \texttt{strategy} class (defaulting to a greedy \texttt{ServerDowntime} breadth-first search).

\textcolor{blue}{\textbf{Trident integration:} Trident edits \texttt{ARTAgent}'s strategy. The generated Python policy (the Coder's output) dynamically replaces the default target selection module via the \texttt{set\_strategy()} method. This architecture allows Trident to dictate high-level tactical routing while relying on the underlying \texttt{ARTAgent} framework to robustly handle the low-level execution of atomic killchain commands, narrowing the generation space and reducing syntax errors.}

\subsubsection{RLVR reward design and specifications}
\label{app:cyberwheel_reward}
To ensure the Trident framework remains versatile and portable across heterogeneous environments, the reward system is decoupled from the simulator's internal engine. Instead, Trident relies on a tactical log-based reward parser, allowing the agent to learn strictly from observable terminal logs and general environment metadata rather than privileged simulator states. Using this log-centric design, rewards are extracted from trajectory text via regular expressions to identify specific tactical events:

\begin{itemize}
    \item \textbf{Positive Milestones} ($\mathcal{E}^{+}$): Rewards are allocated for advancing through the killchain, with weights set at $q_{\text{impact}} = 0.2$, $q_{\text{priv\_esc}} = 0.1$, and $q_{\text{scan}} = 0.05$. To incentivize direct attack paths, an efficiency multiplier ($r = 2.0$) scales the reward based on \autoref{eq:reward}.
    \item \textbf{Negative Events} ($\mathcal{E}^{-}$): To penalize operational noise, inefficiency, and misdirection, the parser identifies failed actions ($s_{\text{fail}} = 0.1$), redundant repetitions ($s_{\text{repeat}} = 0.02$), and interactions with honeypots ($s_{\text{decoy}} = 0.5$). Crucially, treating decoy hits as negative events addresses a misalignment in default simulation reward structures~\citep{cyberwheel}. In CyberWheel's default deception scenario, a red agent receives a positive reward for impacting a decoy, while the blue agent is simultaneously rewarded for a successful misdirection. By classifying decoy interactions as strict negative penalties, we prevent a mutually beneficial local optimum where the red agent naively farms impact rewards on honeypots, ensuring the policy autonomously learns genuine decoy avoidance. The empirical necessity of this penalty to maintain proper adversarial alignment is explicitly demonstrated in our ablation study (\autoref{fig:ablation_decoy} and \autoref{app:decoy}).

\end{itemize}

The final episodic reward is clipped within $R_{\min} = -2.0, R_{\max} = 100.0$ to maintain stability during GRPO gradient optimization.

\subsection{System-level penalties}
\label{app:penalties}

To maintain a stable RLVR training loop and enforce the constraints of the Code-as-Policy paradigm, we define a universal system-level penalty $p$. Aside from environment-specific tactical violation rules, these penalties are shared across both environments to ensure the emergence of robust algorithmic policies. The value of $p$ is determined by the specific failure mode identified during execution:

\begin{equation}
    p = \begin{cases} 
        -10.0 \times N_{\text{leak}} & \text{if } \text{is\_leak}(c) \\
        -10.0 & \text{if } \text{is\_empty}(c) \\
        -5.0  & \text{if } \text{exec\_error}(c) \\
        -2.0  & \text{if } \text{is\_timeout}(c) \\
        -1.0  & \text{if } \text{format\_error}(c) \lor \text{http\_error}(c)
    \end{cases}
\end{equation}

where $N_{\text{leak}}$ denotes the total count of unauthorized access attempts detected in the generated code. The penalty mapping is defined as follows:

\begin{itemize}
    \item \textbf{Information leak }($-10.0 \times N_{\text{leak}}$)\textbf{:} Applied if the policy attempts to bypass the sanitized observation layer to access restricted simulator internals or unobservable ground-truth data (e.g., direct inspection of decoy flags). The penalty scales with the number of detected leak occurrences in the generated code.
    \item \textbf{Empty code ($-10.0$):} A discrete penalty applied if the Coder fails to produce any executable Python logic.
    \item \textbf{Execution error ($-5.0$):} Applied if the generated code triggers unrecoverable runtime errors even after exhausting self-repair attempts.
    \item \textbf{Timeout ($-2.0$):} Applied if the policy execution exceeds the designated time limit, penalizing computationally inefficient or recursive logic.
    \item \textbf{Format \& HTTP error ($-1.0$):} Applied for JSON schema violations or transient infrastructure communication failures. Treating infrastructure-level issues as a minor penalty encourages the agent to prioritize the generation of deployable, robust policies.
\end{itemize}

\section{Heuristic red agent design}
\label{app_b:heuristic}

\subsection{CAGE4 red agent inventory}

Each agent operates within the CybORG CAGE\,4 environment and selects from nine abstract
actions: \texttt{DiscoverRemoteSystems}, \texttt{AggressiveServiceDiscovery},
\texttt{StealthServiceDiscovery}, \texttt{DiscoverDeception}, \texttt{ExploitRemoteService},
\texttt{PrivilegeEscalate}, \texttt{Impact}, \texttt{DegradeServices}, and \texttt{Withdraw}.
Host states are encoded as a nine-symbol alphabet
$\{$\texttt{K}, \texttt{KD}, \texttt{S}, \texttt{SD}, \texttt{U}, \texttt{UD}, \texttt{R}, \texttt{RD}, \texttt{F}$\}$,
representing degrees of knowledge, service discovery, user-level access, root access, and
out-of-reach hosts respectively.

\subsubsection{\texttt{FiniteStateRedAgent} (base class)}

The base finite-state-machine (FSM) red agent maintains a per-host state dictionary and
transitions each host according to separate success and failure state-transition matrices on
every step.
Actions are sampled from a fixed probability matrix conditioned on the current host state,
and decoy services discovered via \texttt{DiscoverDeception} are stored so that
\texttt{ExploitRemoteService} can exclude their PIDs via \texttt{PIDSelectiveExploitActionSelector}.
All derived agents inherit this observation-processing and state-transition logic.

\subsubsection{\texttt{FSRedAgentCombined}}

Extends the base FSM agent with \emph{action-specific} failure tracking: per-host,
per-action-type attempt and failure counts are maintained, and the action-selection
probabilities are dynamically multiplied by $0.1$, $0.5$, or $1.5$ depending on whether the
observed success rate falls below $30\%$, below $50\%$, or exceeds $70\%$, respectively.
Host selection is additionally weighted by overall failure count using the formula
$w = \max(0.1,\,1.5^{-f})$, exponentially suppressing hosts with many failures while still
allowing occasional re-visits.

\subsubsection{\texttt{FSRedAgentDegrader}}

A variant that steers the agent toward service-degradation objectives: hosts already in
state \texttt{R} or \texttt{RD} (root access obtained) receive a $\times 10$ selection weight
boost, and the \texttt{DegradeServices} action probability is multiplied by $4$ at action
selection time regardless of host zone.
The agent otherwise inherits all base FSM state-transition matrices unchanged.

\subsubsection{\texttt{FSRedAgentFailureTracker}}

Tracks \emph{all} action failures per host (not only deception failures) and selects hosts
using exponentially decaying weights $w = \max(0.1,\,1.5^{-f})$, biasing toward hosts with
fewer cumulative failures.
Once a host in state \texttt{S} or \texttt{SD} has accumulated three or more failures,
the \texttt{ExploitRemoteService} probability for that host is zeroed out, shifting focus away
from repeatedly failing exploitation attempts.

\subsubsection{\texttt{FSRedAgentImpacter}}

Prioritises hosts whose hostnames contain \texttt{operational\_zone\_a} or
\texttt{operational\_zone\_b} by assigning them a $\times 100$ selection weight boost.
When acting on a zone-matched host, the \texttt{Impact} action probability is multiplied by
$4$; for unidentified hosts (hostname \texttt{None}), \texttt{DiscoverDeception} is boosted
by $\times 2$; for all other hosts, \texttt{Impact} is suppressed to zero.

\subsubsection{\texttt{FSRedAgentTargetA}}

Targets hosts in \emph{both} \texttt{restricted\_zone\_a} and \texttt{operational\_zone\_a}
with a $\times 100$ selection weight, excluding hosts already in state \texttt{RD} from the
boost to avoid re-exploiting already-owned machines.
Against zone-A hosts, \texttt{ExploitRemoteService} is boosted $\times 4$; for other hosts
\texttt{Withdraw} probability is slightly increased and \texttt{PrivilegeEscalate} halved,
reducing unnecessary escalation outside the target zones.

\subsubsection{\texttt{FSRedAgentTargetAOperational}}

Identical targeting and action-weighting logic to \texttt{FSRedAgentTargetA}, but the zone
keyword is restricted to \texttt{operational\_zone\_a} only, limiting the agent's focus to the
operational (not restricted) subnet of zone A.
Hosts already in state \texttt{RD} within the zone receive a reduced weight of $2$ rather
than $100$, further narrowing aggression to not-yet-fully-owned targets.

\subsubsection{\texttt{FSRedAgentTargetARestricted}}

Mirrors \texttt{FSRedAgentTargetAOperational} but exclusively targets
\texttt{restricted\_zone\_a} hosts, focusing the agent on the restricted (higher-security)
subnet of zone A while leaving operational hosts with default uniform weights.
The same \texttt{RD}-exclusion and action-probability adjustments apply.

\subsubsection{\texttt{FSRedAgentTargetB}}

The zone-B symmetric counterpart of \texttt{FSRedAgentTargetA}: hosts matching
\texttt{restricted\_zone\_b} or \texttt{operational\_zone\_b} receive a $\times 100$ weight
boost (again excluding \texttt{RD} state), and \texttt{ExploitRemoteService} is multiplied by
$4$ when acting on zone-B hosts.
Outside zone B, \texttt{PrivilegeEscalate} is halved and \texttt{Withdraw} is marginally
increased ($+0.1$).

\subsubsection{\texttt{FSRedAgentTargetBOperational}}

Focuses on \texttt{operational\_zone\_b} hosts only, following the same $\times 100$ weight
and $\times 4$ exploit-boost pattern as \texttt{FSRedAgentTargetB}, but confined to the
operational subnet.
Hosts already at \texttt{RD} are not excluded from the weight boost here (unlike the zone-A
operational variant), so the agent continues to visit fully-compromised operational-B hosts.

\subsubsection{\texttt{FSRedAgentTargetBRestricted}}

Mirrors \texttt{FSRedAgentTargetBOperational} but scoped exclusively to
\texttt{restricted\_zone\_b}, directing exploitation pressure solely at the restricted subnet
of zone B.
Action-weighting is identical to the other zone-B variants: exploit boost on target hosts,
reduced escalation and slightly raised withdrawal elsewhere.

\subsubsection{\texttt{RandomSelectRedAgent}}

A baseline agent that uniformly samples a random valid action from the full action space
at each step, with no host-state tracking or FSM logic.
Parameter values (IP addresses, hostnames, subnets) are also chosen uniformly at random
from the valid options in the current action space; an optional \texttt{disable\_withdraw}
flag can remove the \texttt{Withdraw} action from consideration.

\subsection{CyberWheel red strategy inventory}

Each class in this directory implements a \emph{target-selection strategy} for the CyberWheel
red agent.
All strategies inherit from the abstract base class \texttt{RedStrategy} and override a single
class method, \texttt{select\_target(agent\_obj) -> Host}, which returns the host the red
agent should act on in the current step.
The agent's internal history (\texttt{agent\_obj.history}), known unimpacted servers
(\texttt{agent\_obj.unimpacted\_servers}), and the set of unknown hosts
(\texttt{agent\_obj.unknowns}) are the primary inputs to every strategy's decision.
The difference between the plain and BFS-prefixed variants lies solely in traversal order:
plain strategies draw the next host \emph{randomly} from a pool, whereas BFS strategies
take the \emph{front element} (\texttt{data\_list[0]}) for deterministic breadth-first
exploration.


\subsubsection{\texttt{BruteForce}}

The simplest possible strategy: \texttt{select\_target} unconditionally returns
\texttt{agent\_obj.current\_host}, meaning the agent repeatedly attacks the host it is
already on without ever moving.
This serves as a lower-bound baseline and is useful for stress-testing a single host's
defences or verifying killchain execution in isolation.

\subsubsection{\texttt{Exfiltration}}

Designed to locate and impact a designated \emph{leader} (target) host.
The agent stays on the current host if it is still \texttt{Unknown} (i.e.\ not yet fully
characterised) or if it has been identified as the leader; otherwise it picks a
\emph{random} unknown host from \texttt{agent\_obj.unknowns}, falling back to the current
host if the pool is empty.
Stale names that no longer appear in the network graph are silently pruned from the unknown
pool during selection.

\subsubsection{\texttt{ServerDowntime}}

Aims to impact every server reachable in the network, one by one.
The agent remains on the current host if it is \texttt{Unknown} or is still listed in
\texttt{agent\_obj.unimpacted\_servers}; otherwise it draws a \emph{random} host from the
unimpacted-server pool, then falls back to the unknown-host pool, and finally stays put if
both are empty.
Stale hostnames are pruned from whichever pool is being consumed when they are not found in
the live network graph.

\subsubsection{\texttt{DFSImpact}}

Implements a depth-first impact sweep: the agent attacks the current host until it has
completed the full killchain (i.e.\ \texttt{last\_step == len(killchain) - 1}), then
immediately jumps to a \emph{random} unimpacted host drawn from
\texttt{agent\_obj.unimpacted\_hosts}.
Because no breadth-first bookkeeping is performed, the traversal order across hosts is
non-deterministic, giving the strategy a randomised DFS character.

\subsubsection{\texttt{BFSExfiltration}}

A breadth-first variant of \texttt{Exfiltration}: the agent stays on the current host if it
is \texttt{Unknown} or the leader, but when moving it selects
\texttt{agent\_obj.unknowns.data\_list[0]} — the oldest-inserted entry — instead of a
random one.
This FIFO access pattern ensures unknown hosts are explored in discovery order, producing a
systematic breadth-first sweep toward the target leader host.

\subsubsection{\texttt{BFSServerDowntime} \textbf{(base class)}} 

A breadth-first variant of \texttt{ServerDowntime}: the agent stays on the current host if
it is \texttt{Unknown} or unimpacted, then moves to
\texttt{unimpacted\_servers.data\_list[0]} (oldest known unimpacted server) if available,
or \texttt{unknowns.data\_list[0]} otherwise.
The deterministic FIFO ordering guarantees that servers are impacted in the order they were
first discovered, making this strategy more predictable and reproducible than the randomised
\texttt{ServerDowntime}.
\section{Ablation study}
\label{app:c_reward_ablation}

\subsection{Cross-environment validation: reward correlations}

To validate the efficacy of Trident's reward design, we evaluate the reward formulation at a macro level by analyzing the relationship between the red agent's verifiable red reward and the actual degradation of the defender's blue reward.

\begin{figure}[!h]
    \centering
    \includegraphics[width=\linewidth]{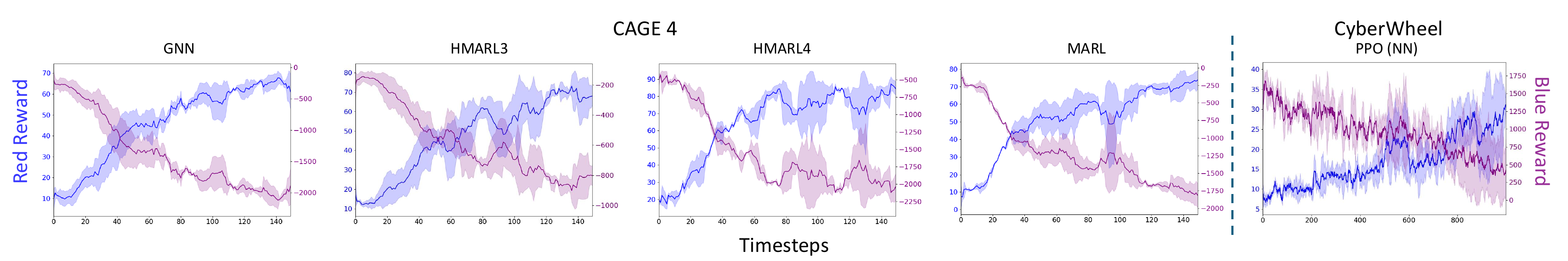}
    \caption{\textbf{Learning dynamics of verifiable tactical rewards against defender performance.} As the RLVR process progresses (x-axis: episodes), the red agent successfully maximizes its verifiable reward, consistently driving the blue agent's accumulated reward into a steep decline.}
    \label{fig:reward_corr}
\end{figure}

\begin{figure}[!h]
    \centering
    \includegraphics[width=\linewidth]{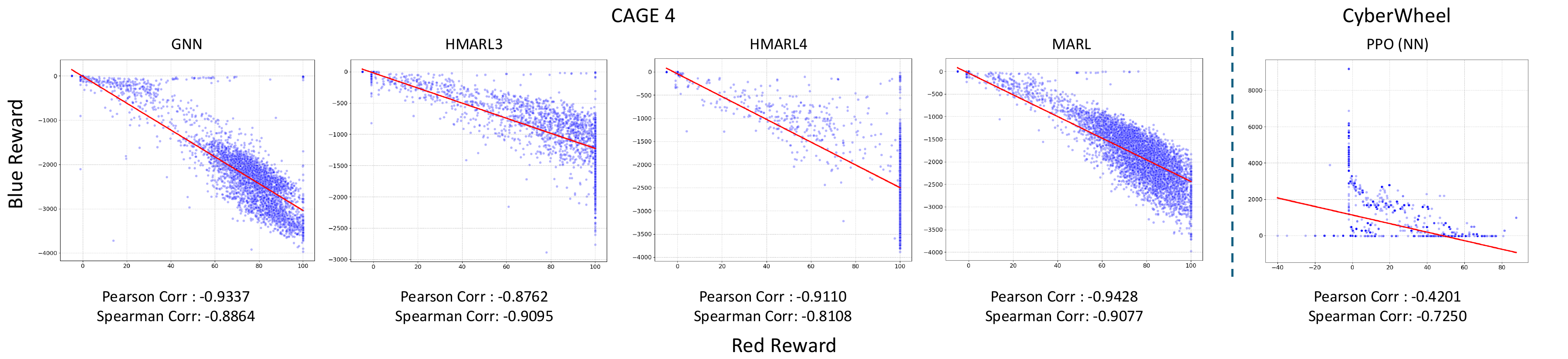}
    \caption{\textbf{Correlation between verifiable tactical rewards and defender performance.} Scatter plots illustrating the strong negative correlation between Trident's episodic rewards and the blue agent's rewards. The high correlation coefficients validate that our verifiable reward signal serves as a high-fidelity proxy for actual system compromise across different environments.}
    \label{fig:reward_scatter}
\end{figure}

This relationship is demonstrated through two distinct perspectives. First, \autoref{fig:reward_corr} illustrates the learning dynamics over time. Across all evaluated DRL defenders in both CAGE 4 and CyberWheel, as the red agent optimizes its policy and the Red Reward increases, the Blue Reward experiences a mirrored, consistent degradation. Second, \autoref{fig:reward_scatter} formalizes this relationship statistically via scatter plots of episodic outcomes, removing the temporal dimension. In CAGE 4, the Pearson correlation coefficients range from $-0.87$ to $-0.94$, demonstrating an exceptionally tight mathematical alignment between our rule-based proxy rewards and the true systemic compromise of the DRL defenders. A similar negative trend is successfully maintained in CyberWheel (Spearman: $-0.73$), confirming the cross-environment robustness of the reward framework.

\textbf{Note on micro-level ablation:} While a fine-grained ablation of individual tactical weights (e.g., $q_{\text{impact}}$ vs. $q_{\text{priv\_esc}}$) would be ideal, exhaustively tuning these parameters presents a combinatorial explosion that is practically infeasible. Therefore, we rely on heuristic weights based on domain expertise for this study, leaving automated reward optimization as future work.

\subsection{Architectural and performance ablations in CAGE 4}

To isolate the impact of Trident Agentic's core architectural and reward design choices, we conducted further ablation studies within the CAGE 4 environment. For the victim defender, we specifically selected the \textbf{HMARL-3} agent since HMARL-3 demonstrated the most consistent learning stability and performance that can provide a reliable and robust baseline for sensitivity analysis. We evaluated two degraded configurations against our full framework: removing the efficiency multiplier ($w/o \text{ efficiency bonus}$, $r=0$) and disabling the Coder's self-repair mechanism ($w/o \text{ retry, } K=1$). The results, averaged over three independent training runs, are summarized in \autoref{tab:ablation}.

\begin{table}[h!]
    \centering
    \caption{\textbf{Ablation study of Trident's architectural and reward components in CAGE 4.} Lower mean values indicate stronger red agent performance (greater degradation of the blue agent). Displaying individual seeds (Run 1--3) reveals that while $K=1$ achieves a lower overall mean, it suffers from critical instability driven by a single outlier, whereas $K=3$ remains consistently robust.}
    \label{tab:ablation}
    \begin{tabular}{lcccc}
        \toprule
        \textbf{Ablation Configuration} & \textbf{Run 1} & \textbf{Run 2} & \textbf{Run 3} & \textbf{Mean $\pm$ Std. Dev. $\downarrow$} \\
        \midrule
        Trident (Ours, $K=3$) & -843.43 & -856.35 & -849.76 & -849.85 $\pm$ \textbf{6.46} \\
        w/o Efficiency Bonus ($r=0$)  & -884.68 & -855.80 & -767.61 & -836.03 $\pm$ 60.99 \\
        w/o Retry Loop ($K=1$) & -646.04 & -1358.14 & -741.14 & \textbf{-915.11} $\pm$ 386.62 \\
        \bottomrule
    \end{tabular}
\end{table}

\textbf{The necessity of the self-repair Loop ($K=1$):} Restricting the Coder to a single generation attempt ($K=1$) without the AST-based self-repair loop drastically destabilizes the RLVR training process. Although the $K=1$ configuration yields a lower mean reward, this metric is deceptive; it is entirely driven by a single highly successful outlier run, masking an extreme generation variance. Without the ability to absorb and correct minor syntactic errors or transient hallucinations, the GRPO signal becomes dominated by execution penalties. This confirms that the $K=3$ self-repair loop is indispensable for filtering out syntax noise, ensuring that the Planner learns a consistently robust and reproducible strategy.

\textbf{The role of the efficiency bonus:} Removing the efficiency multiplier ($r=0$ in \autoref{eq:reward}) from the tactical reward signal results in a slightly degraded mean performance and an order-of-magnitude increase in variance. While the agent can still learn effective attack paths without explicit incentives for rapid execution, the absence of this bonus makes the reward signal less focused, leading to higher instability across individual seeds. This indicates that the efficiency bonus contributes positively to both the overall maximization of the reward and the stability of the policy optimization.

\subsection{Environment-specific alignment: decoy penalties in CyberWheel}
\label{app:decoy}
While the macro-level correlations demonstrate general robustness, adapting RLVR to specific environments often requires handling domain-specific simulation mechanics. To validate our handling of such mechanics, we ablate the decoy penalty ($s_{\text{decoy}}$) within the CyberWheel environment. 

\begin{figure}[!h]
    \centering
    \includegraphics[width=\linewidth]{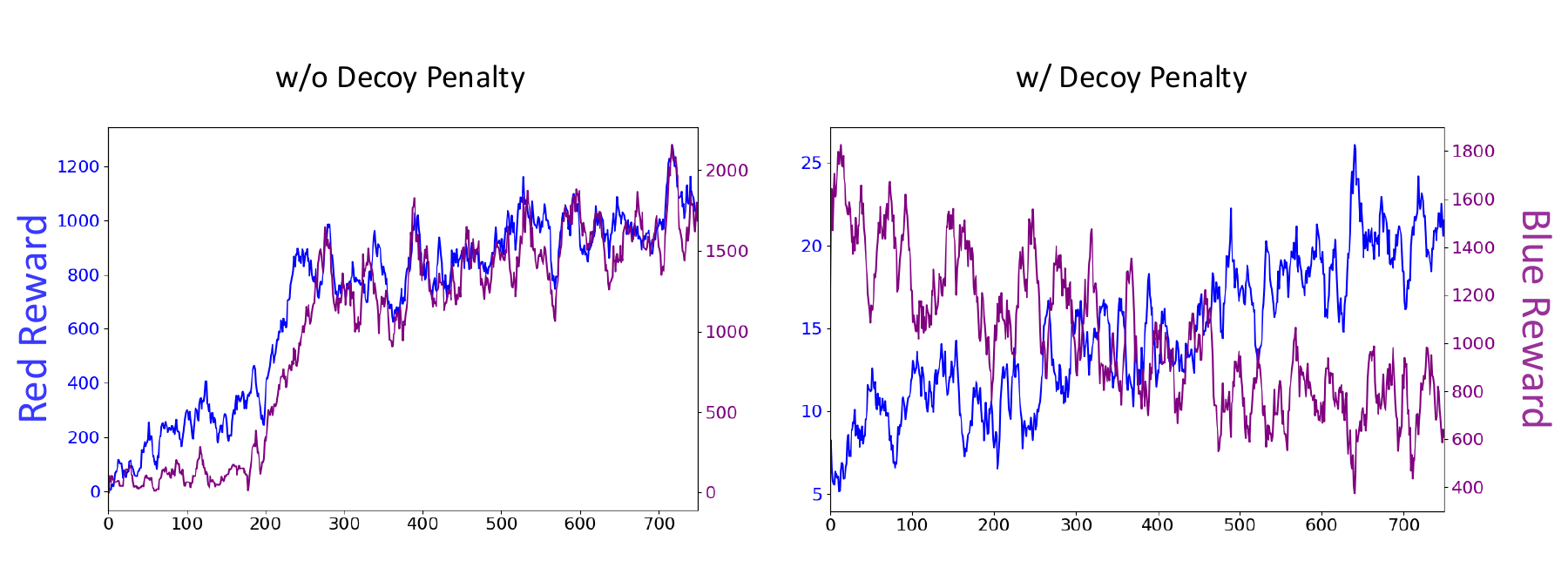}
    \vspace{-5mm}
    \caption{\textbf{Impact of decoy interaction penalties on optimization alignment in CyberWheel.} (Left) Without a specific penalty, the agent reaches a misaligned local optimum by repeatedly interacting with decoys, mutually benefiting both agents. (Right) By incorporating a strict negative penalty for decoy hits, Trident learns to avoid honeypots, effectively reducing the blue agent's reward while minimizing operational noise.}
    \label{fig:ablation_decoy}
\end{figure}

As illustrated in \autoref{fig:ablation_decoy}, the presence or absence of this penalty fundamentally alters the agent's learned behavior. When no explicit penalty is applied for interacting with honeypots (left), the Red Reward and Blue Reward paradoxically increase together. This exposes a critical misalignment: the red agent discovers a local optimum by farming impact rewards on decoys, while the blue defender is rewarded for successful misdirection. By enforcing a strict decoy penalty (right), the red agent is forced to autonomously learn behavioral patterns to identify and avoid honeypots. The absolute scale of the Red Reward is significantly lower compared to the unpenalized configuration, reflecting the transition from infinite decoy farming to precise, targeted exploitation of genuine servers.

\section{Simulator-Native PPO Baseline}
\label{app:ppo_baseline}

\noindent\textbf{Implementation Details.}
We implement the DRL red-agent baseline using a hand-written PPO
implementation without Stable-Baselines3. A separate PPO agent is trained
against each frozen blue defender. The policy and value functions use
separate two-layer MLPs with 64 hidden units per layer, ReLU activations,
and orthogonal initialization. Training uses generalized advantage
estimation (GAE), the clipped PPO objective, entropy regularization, and
value-function learning.

For CAGE4, \texttt{DRLRedAgent} subclasses
\texttt{FiniteStateRedAgent} and retains its host-state tracking and
action-parameter resolution. PPO replaces its rule-based host and action
selection logic. The wrapper converts the dynamically generated network and
the red agent's current host knowledge into a fixed-length numerical
observation and flattens the available action--host combinations into a
discrete action space with an additional no-op action. Invalid action--host
combinations fall back to \texttt{Sleep()} and receive a small penalty; no
hard action masking is applied.

For CyberWheel, the PPO attacker is trained against the frozen PPO blue
defender through the simulator-native red-agent interface. In both
environments, PPO therefore learns a step-wise numerical policy directly
from simulator transitions, without the Log Summarizer, Planner, Coder, or
GRPO components used by Trident Agentic.

\noindent\textbf{Training Budget and Evaluation Protocol.}
Each PPO red agent is trained for 4 million simulator steps against its
corresponding frozen defender. Training trajectories are generated in
parallel on CPU using \texttt{joblib.Parallel} with process-based execution.

For CAGE4, Trident Agentic at the reported training step 150 uses two
contexts per optimization step, $G=8$ rollouts per context, and 500 simulator
steps per rollout, corresponding to
\[
150 \times 2 \times 8 \times 500 = 1.2 \text{ million}
\]
simulator steps. Thus, the PPO baseline receives approximately $3.33\times$
more simulator interaction than Trident Agentic at the reported checkpoint,
reducing the likelihood that weaker PPO performance is caused by insufficient
environment exposure.

Both methods are evaluated under the same CAGE4 protocol: 500 episodes of
500 simulator steps, repeated across three independent runs. Exact
computational matching is not meaningful because PPO and Trident differ in
model size, observation representation, policy parameterization, update
frequency, and problem formulation. PPO learns a step-wise policy from
numerical transitions, whereas Trident generates a complete executable policy
and receives episode-level feedback.
\section{Dataset Analysis}
\label{app:f_dataset}

We characterize the Trident trajectory dataset $\mathcal{D}$ in terms of
coverage, diversity, outcome distribution, redundancy, and train--validation
consistency. The released dataset contains 9,500 training and 500 validation
trajectories for CAGE~4, and 3,556 training and 188 validation trajectories for
CyberWheel. \autoref{tab:dataset_summary} summarizes the main statistics.

\subsection{Dataset Statistics}
\label{app:f_schema}

Each trajectory contains a serialized execution log (\texttt{raw\_log}), the
source heuristic code (\texttt{current\_code}), the generating heuristic label,
and, for CyberWheel, the episodic reward. Since \texttt{current\_code} is fixed
within each heuristic group, trajectory-level statistics are computed from
\texttt{raw\_log}.

\begin{table}[!h]
    \centering
    \caption{\textbf{Summary of the released Trident dataset.}
    Action-type sequences are extracted from \texttt{raw\_log} after removing
    hostnames, IP addresses, and step indices.}
    \label{tab:dataset_summary}
    \resizebox{\columnwidth}{!}{
    \begin{tabular}{lcc}
        \toprule
         & \textbf{CyberWheel} & \textbf{CAGE 4} \\
        \midrule
        Train / validation trajectories & 3,556 / 188 & 9,500 / 500 \\
        Source generators & 5 strategies & 12 agents \\
        Trajectories per generator (train) & 672--765 & 747--852 \\
        \texttt{raw\_log} length (chars), mean $\pm$ std
            & $18{,}761 \pm 9{,}931$ & $110{,}814 \pm 44{,}377$ \\
        Documented action vocabulary observed & 6 / 6 & 9 / 9 \\
        Action unigram / bigram entropy (bits) & 2.49 / 3.57 & 2.96 / 5.87 \\
        Unique action-type sequences
            & 2,570 / 3,556 (72.3\%) & 8,546 / 9,500 (90.0\%) \\
        Exact \texttt{raw\_log} duplicates & 0 & 0 \\
        Train--val consistency ($p$-values)
            & $\chi^2$ 0.155; KS length 0.733; KS reward 0.586
            & $\chi^2$ 0.667; KS length 0.251 \\
        \bottomrule
    \end{tabular}
    }
\end{table}

\subsection{Coverage}
\label{app:f_coverage}

\begin{figure}[!h]
    \centering
    \includegraphics[width=\linewidth]{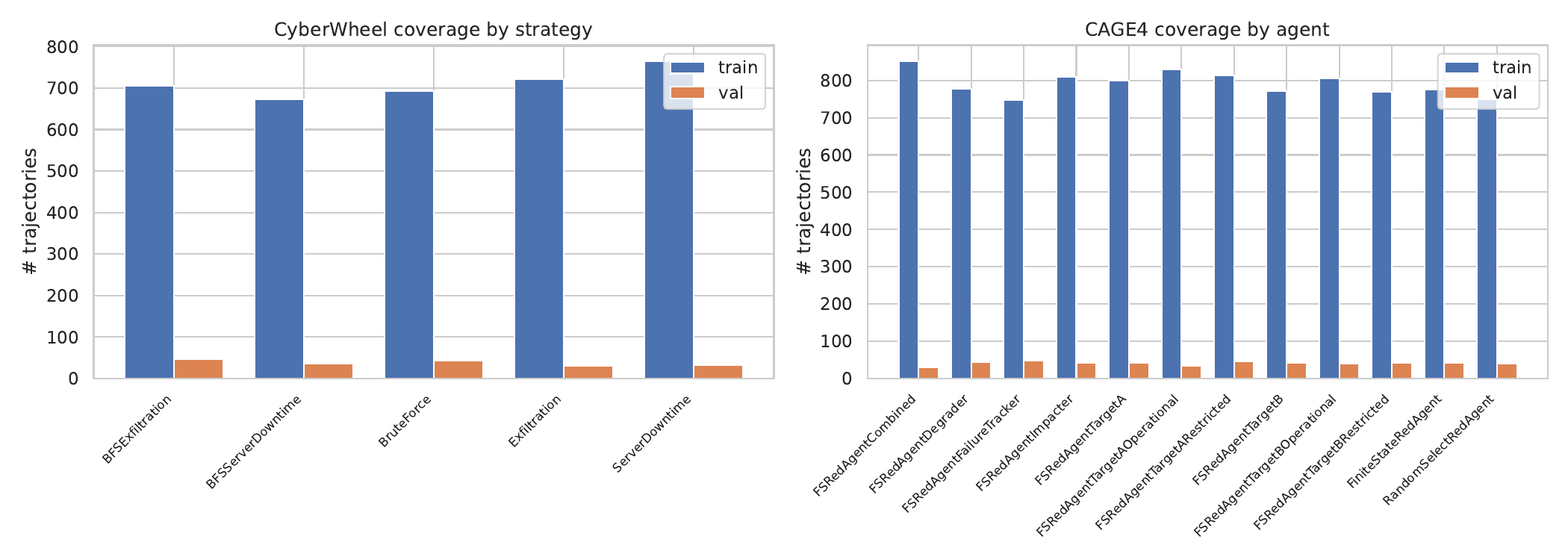}
    \caption{\textbf{Coverage by generating heuristic.}
    Number of training and validation trajectories generated by each CyberWheel
    strategy (left) and CAGE~4 agent (right).}
    \label{fig:dataset_coverage}
\end{figure}

Trajectories are distributed approximately uniformly across the generating
heuristics (\autoref{fig:dataset_coverage}). Training counts range from
672--765 across the five CyberWheel strategies and 747--852 across the twelve
CAGE~4 agents, with a maximum-to-minimum ratio of 1.14 in both environments.

\subsection{Diversity}
\label{app:f_diversity}

The dataset covers all documented strategic actions in both environments
(\autoref{tab:dataset_vocab}). Action unigram / bigram entropy is
2.49 / 3.57 bits for CyberWheel and 2.96 / 5.87 bits for CAGE~4.
Moreover, 72.3\% of CyberWheel and 90.0\% of CAGE~4 trajectories have unique
action-type sequences, with no exact \texttt{raw\_log} duplicates.

\begin{table}[!h]
    \centering
    \caption{\textbf{Action types observed in the training split.}}
    \label{tab:dataset_vocab}
    \small
    \begin{tabular}{lp{0.72\linewidth}}
        \toprule
        \textbf{Environment} & \textbf{Observed action types} \\
        \midrule
        CyberWheel (6 / 6) &
        \texttt{pingsweep}, \texttt{portscan}, \texttt{discovery},
        \texttt{lateral-movement}, \texttt{privilege-escalation},
        \texttt{impact} \\
        \midrule
        CAGE 4 (9 / 9) &
        \texttt{DiscoverRemoteSystems}, \texttt{AggressiveServiceDiscovery},
        \texttt{StealthServiceDiscovery}, \texttt{DiscoverDeception},
        \texttt{ExploitRemoteService}, \texttt{PrivilegeEscalate},
        \texttt{Impact}, \texttt{DegradeServices}, \texttt{Withdraw} \\
        CAGE 4 (auxiliary) &
        \texttt{Sleep}, \texttt{RedSessionCheck} \\
        \bottomrule
    \end{tabular}
\end{table}

\begin{figure}[!h]
    \centering
    \includegraphics[width=\linewidth]{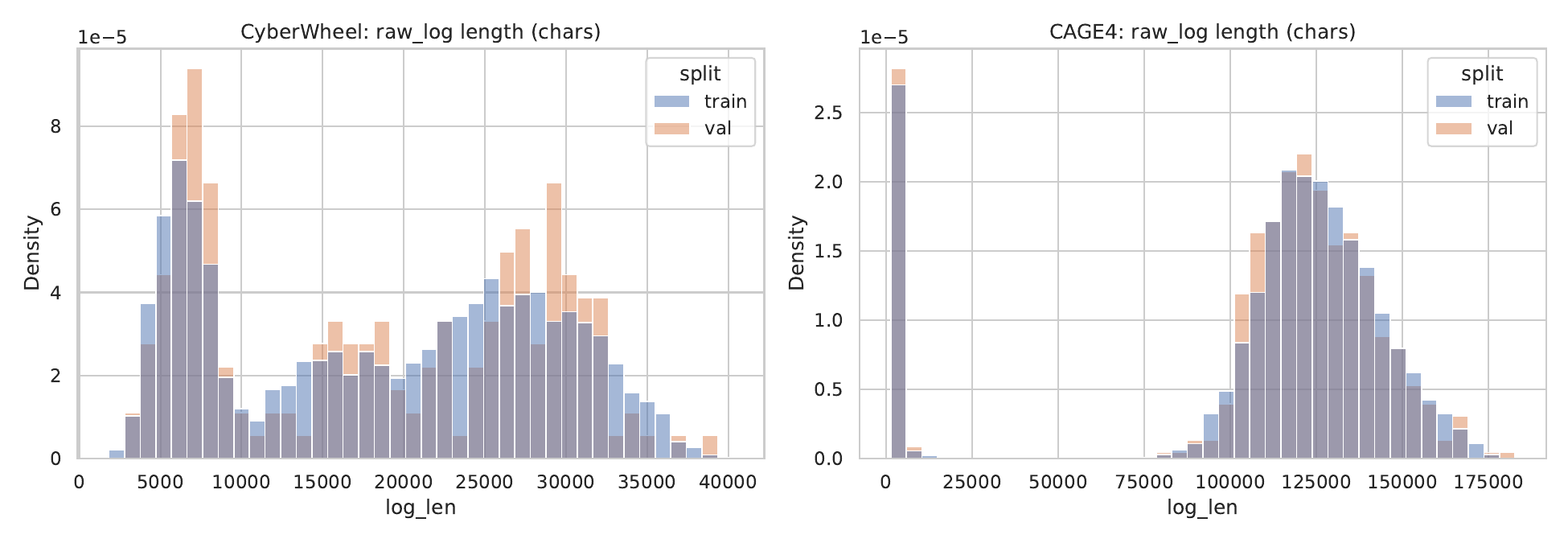}
    \caption{\textbf{Log-length distribution.}
    Distribution of \texttt{raw\_log} length for the training and validation
    splits of CyberWheel (left) and CAGE~4 (right).}
    \label{fig:dataset_diversity}
\end{figure}

\autoref{fig:dataset_diversity} further shows substantial variation in
serialized trajectory length, reflecting differences in episode progression
across randomized environments and defender interactions.

\subsection{Outcome Distribution}
\label{app:f_outcome}

\begin{figure}[!h]
    \centering
    \includegraphics[width=\linewidth]{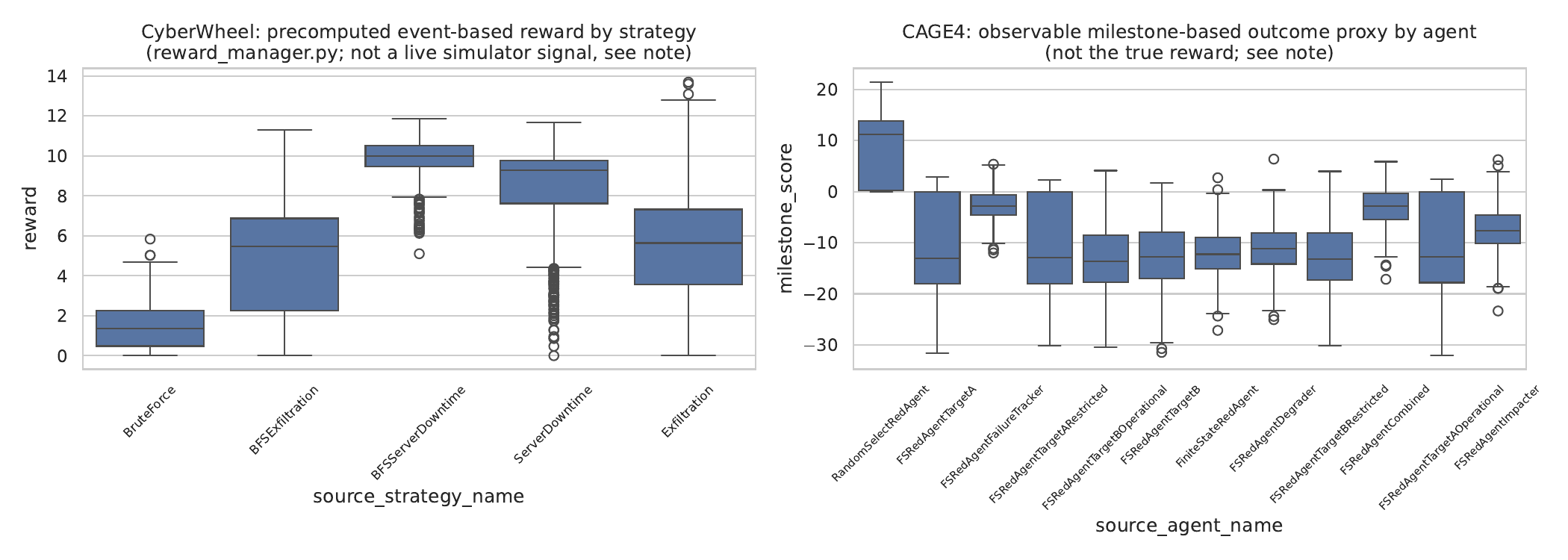}
    \caption{\textbf{Outcome distribution by generating heuristic.}
    CyberWheel episodic reward (left) and CAGE~4 observable milestone score
    (right).}
    \label{fig:dataset_difficulty}
\end{figure}

\autoref{fig:dataset_difficulty} shows that the generating heuristics produce a
broad range of episode outcomes. CyberWheel uses the released episodic reward.
Because CAGE~4 trajectories do not contain a reward field, we use an observable
milestone score computed from successful and failed actions in
\texttt{raw\_log}. The CAGE~4 score is used only for relative comparison among
generating heuristics.

\subsection{Redundancy}
\label{app:f_redundancy}

\begin{figure}[!h]
    \centering
    \includegraphics[width=\linewidth]{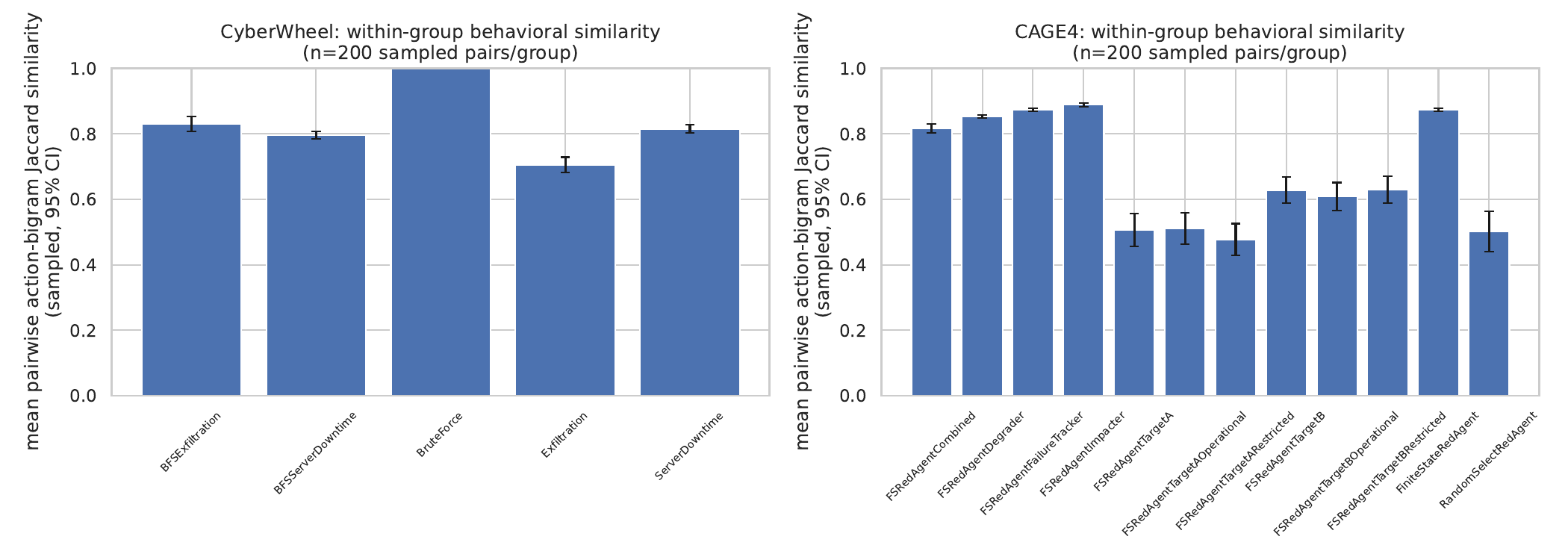}
    \caption{\textbf{Within-generator behavioral similarity.}
    Mean pairwise Jaccard similarity of action-type bigram sets over
    200 sampled trajectory pairs per generator.}
    \label{fig:dataset_redundancy}
\end{figure}

We measure within-generator redundancy using the Jaccard similarity of
action-type bigram sets over 200 sampled trajectory pairs per generator
(\autoref{fig:dataset_redundancy}). Deterministic heuristics show higher
similarity, while heuristics affected by randomized network states and defender
responses exhibit greater variation.

Action-sequence similarity is a conservative measure of log similarity:
\texttt{raw\_log} also contains hostnames, IP addresses, discovered services,
and environment-specific state information. The training split contains
986 CyberWheel and 954 CAGE~4 action-sequence collisions, but no exact
\texttt{raw\_log} duplicates.

\subsection{Train--Validation Consistency}
\label{app:f_consistency}

\begin{figure}[!h]
    \centering
    \includegraphics[width=0.8\linewidth]{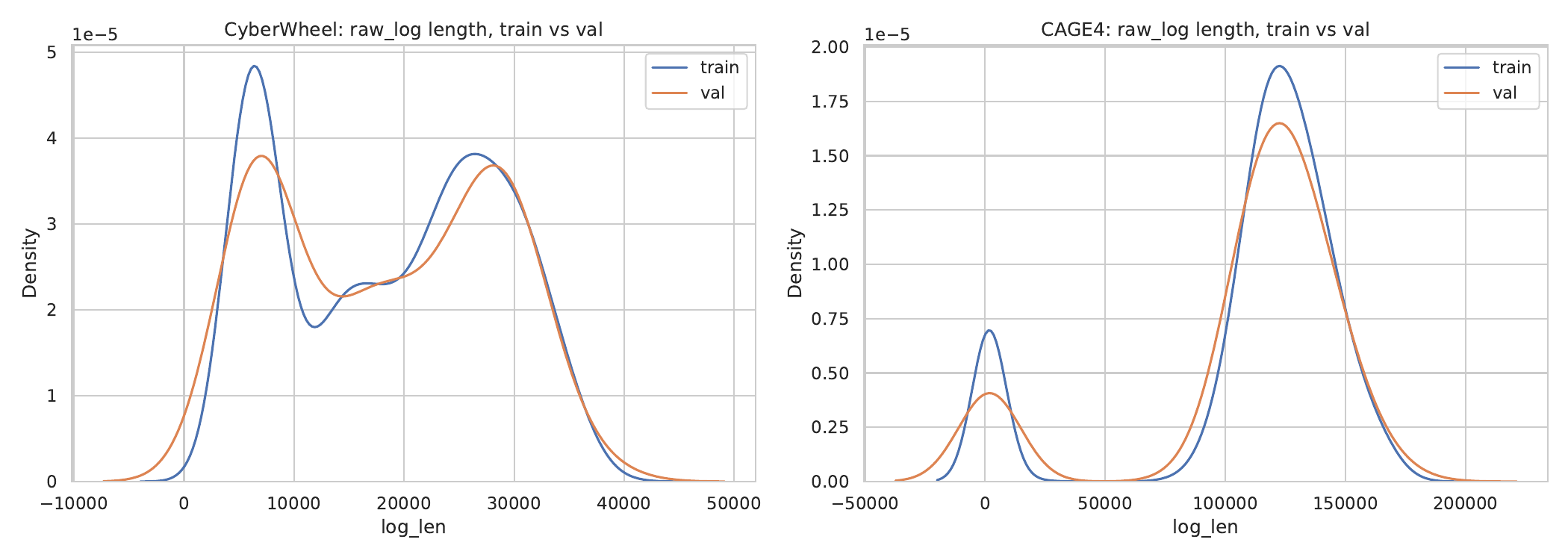}
    \caption{\textbf{Training and validation log-length distributions.}
    Kernel density estimates for CyberWheel (left) and CAGE~4 (right).}
    \label{fig:dataset_consistency}
\end{figure}

We compare the training and validation splits using a $\chi^2$ test on
generator proportions and two-sample KS tests on log length and CyberWheel
reward. All tests yield $p>0.15$ (\autoref{tab:dataset_consistency}), with
closely overlapping log-length distributions
(\autoref{fig:dataset_consistency}).

\begin{table}[!h]
    \centering
    \caption{\textbf{Train--validation consistency tests.}}
    \label{tab:dataset_consistency}
    \begin{tabular}{llcc}
        \toprule
        \textbf{Environment} & \textbf{Test} & \textbf{Statistic} & \textbf{$p$-value} \\
        \midrule
        CyberWheel & Strategy proportions ($\chi^2$, df $=4$) & 6.67 & 0.155 \\
        CyberWheel & \texttt{raw\_log} length (KS) & 0.050 & 0.733 \\
        CyberWheel & Reward (KS) & 0.057 & 0.586 \\
        CAGE 4 & Agent proportions ($\chi^2$, df $=11$) & 8.51 & 0.667 \\
        CAGE 4 & \texttt{raw\_log} length (KS) & 0.046 & 0.251 \\
        \bottomrule
    \end{tabular}
\end{table}

\end{document}